\documentclass[runningheads]{llncs}
\pdfoutput=1

\usepackage{microtype}
\usepackage{preamble}
\usepackage[hidelinks]{hyperref}
\usepackage{mdframed}
\usepackage{tikz}
\usepackage{cleveref}
\usepackage{graphicx}
\usepackage[obeyFinal]{todonotes}
\usepackage{xspace}
\usepackage{subcaption}
\usepackage{xcolor}
\usepackage{nccmath}
\usepackage{booktabs}
\usepackage{enumitem}
\usepackage{longtable}
\usepackage{xfrac}
\usepackage{adjustbox}
\usepackage{thmtools}
\usepackage{orcidlink}
\usepackage{comment}

\newcommand{\appcite}[2]{#2} 

\newcommand{\indicator}[1]{[#1]}

\newcommand{\Distr}{\Delta}
\newcommand{\Obs}{Z}
\newcommand{\act}{a}

\newcommand{\actps}[3]{\left\langle #1, \langle #2, #3\rangle \right\rangle}

\newcommand{\talrm}{t_{\text{alrm}}}
\newcommand{\tsafe}{t_{\text{safe}}}
\newcommand{\tig}{t_{\text{ignr}}}

\newcommand{\icy}{{\color{cyan} icy}\xspace}
\newcommand{\dry}{{\color{brown} dry}\xspace}
\newcommand{\ig}{z_{\color{gray} ignore}\xspace}
\newcommand{\en}{z_{\color{black} end}\xspace}

\newcommand{\alarmas}[1][h+1]{\left\langle #1, \atop \talrm\right\rangle}
\newcommand{\safeas}[1][h+1]{\left\langle #1, \atop \tsafe\right\rangle}
\newcommand{\igas}[1][h+1]{\left\langle #1, \atop \tig\right\rangle}
\newcommand{\alarms}[1][h+1]{\left\langle #1, \talrm\right\rangle}
\newcommand{\safes}[1][h+1]{\left\langle #1, \tsafe\right\rangle}
\newcommand{\igs}[1][h+1]{\left\langle #1, \tig\right\rangle}

\newcommand{\safetraces}[2]{\mathbb{S}_{#1}^{#2}}
\newcommand{\unsafetraces}[2]{\mathbb{U}_{#1}^{#2}}
\newcommand{\missedalarms}[2]{\mathsf{mA}_{#1}^{#2}}
\newcommand{\falsealarms}[2]{\mathsf{fA}_{#1}^{#2}}
\newcommand{\defeq}{:=}

\newcommand{\ccSched}{\Sigma_c}

\newcommand{\EQ}{\operatorname{EQ}}
\newcommand{\MQ}{\operatorname{MQ}}

\renewcommand{\paragraph}[1]{\smallskip\noindent\emph{#1}}
\spnewtheorem*{proofoutline}{Proof Sketch}{\itshape}{\rmfamily}
\usetikzlibrary{automata, positioning, arrows.meta, fit}
\tikzset{
    >=stealth, 
    every state/.style={thick, fill=gray!10, minimum size=0.6cm,inner sep=1pt}, 
    initial text=$ $, 
    action/.style={circle, fill, inner sep=1pt},
    icy/.style={draw=cyan},
    dry/.style={draw=brown},
    return/.style={color=lightgray, line width=1mm},
    other/.style={rounded rectangle},
    problem/.style={rectangle, draw, minimum height=1cm, minimum width=1.5cm, align=center}
}
\usetikzlibrary{patterns, intersections, positioning, shapes.geometric, shapes.misc, shapes.symbols, calc,
	automata, arrows, decorations.pathreplacing,decorations.pathmorphing, backgrounds}

\crefname{figure}{Fig.}{Figs.}
\Crefname{figure}{Figure}{Figures}
\crefname{section}{Sec.}{Secs.}
\Crefname{section}{Section}{Sections}
\crefname{subsection}{Sec.}{Secs.}
\Crefname{subsection}{Section}{Sections}
\crefname{definition}{Def.}{Defs.}
\Crefname{definition}{Definition}{Definitions}
\crefname{theorem}{Thm.}{Thms.}
\Crefname{theorem}{Theorem}{Theorems}
\crefname{lemma}{Lem.}{Lems.}
\Crefname{lemma}{Lemma}{Lemmas}
\crefname{enumi}{Prob.}{Probs.}
\Crefname{enumi}{Problem}{Problems}

\crefname{appendix}{App.}{Apps.}
\Crefname{appendix}{Appendix}{Appendices}
\crefname{appsec}{App.}{Apps.}
\Crefname{appsec}{Appendix}{Appendices}

\title{Learning Verified Monitors for Hidden~Markov~Models\thanks{This work has been partially funded by the NWO grant FuRoRe (OCENW.M.22.282).}}

\author{Luko van der Maas \and Sebastian Junges }
\institute{Radboud University, Nijmegen, the Netherlands}
\author{%
 Luko~van~der~Maas \orcidlink{0009-0007-3915-6191} \and Sebastian~Junges \orcidlink{0000-0003-0978-8466}
}
\institute{Radboud University, Nijmegen, the Netherlands\\ \email{\{luko.vandermaas,sebastian.junges\}@ru.nl}}

\newcommand{\alg}{ToVer\xspace}

\begin{document}
\maketitle
\begin{abstract}
    Runtime monitors assess whether a system is in an unsafe state based on a stream of observations. We study the problem where the system is subject to probabilistic uncertainty and described by a hidden Markov model. A stream of observations is then unsafe if the probability of being in an unsafe state is above a threshold. A correct monitor recognizes the set of unsafe observations. The key contribution of this paper is the first correct-by-construction synthesis method for such monitors, represented as finite automata. The contribution combines four ingredients: First, we establish the coNP-hardness of checking whether an automaton is a correct monitor, i.e., a monitor without misclassifications. Second, we provide a reduction that reformulates the search for misclassifications into a standard probabilistic system synthesis problem. Third, we integrate the verification routine into an active automata learning routine to synthesize correct monitors. Fourth, we provide a prototypical implementation that shows the feasibility and limitations of the approach on a series of benchmarks.
\end{abstract}

\section{Introduction}
\noindent
Runtime assurance is an essential ingredient in the deployment of safe autonomous systems~\cite{bartocciLecturesRuntimeVerification2018,sanchezSurveyChallengesRuntime2019}.
Runtime monitors provide assurance by flagging potentially dangerous system behavior, based on a system execution.
More precisely, a monitor receives a stream of observations about the system and outputs a verdict, e.g., it raises an alarm that the system has left some safety envelope. A monitor is correct if it correctly raises such alarms based on a formal specification.
Various challenges in creating correct runtime monitors for (semi-)autonomous systems have been identified~\cite{sanchezSurveyChallengesRuntime2019}, such as:
(1)~the state of the system is only partially observable, i.e., the stream of observations comes from sensor readings and does not uniquely identify the state of a system,
(2)~the behavior of the system and/or the sensors may be subject to probabilistic uncertainty,
(3)~the monitor itself is subject to resource constraints (time, memory, etc.), and
(4)~the monitor is itself safety-critical and should therefore be subject to extensive validation.
Challenges (1,2) can be addressed  by modelling the system as a hidden Markov model (HMM), and Challenges (3,4) can be addressed by representing a monitor as, e.g., a (small) finite automaton.
Concretely, this paper focuses on the following question:
\emph{Is a given finite automaton a correct monitor for a given and known HMM?}
This paper studies the complexity of this problem, provides a practical verification approach, and embeds it into a framework to learn monitors.

\paragraph{What are HMMs?}
In this paper, we assume that the system including its sensors is adequately modeled as a discrete Hidden Markov Model (HMM)~\cite{rabinerTutorialHiddenMarkov1989} and that we have full access to this HMM.
Markov chains (MCs) describe system behavior subject to probabilistic uncertainty.
Paths through an MC are sequences of states that describe system executions.
HMMs extend MCs by labelling their states with \emph{observations}.
Intuitively, the observations can be used to model the information that the monitor receives in every state.
In HMMs, every path can be lifted to a sequence of observations, which we call a \emph{trace}.
The trace associated to a system execution describes the information received by the monitor.

\begin{figure}[t]
    \vspace{-3mm}
    \begin{minipage}{.49\textwidth}
        \centering

        \tikzset{
            pics/alarm/.style args={#1}{
                    code={
                            \draw[fill=black] (0,0) rectangle (3,0.3);
                            \draw[rounded corners, fill=black] (0.2,0.2) rectangle (2.8,0.5);
                            \draw[fill=#1] (0.5,0.5) -- (0.5,1.5) to[out=90,in=90] (2.5,1.5) -- (2.5,0.5) -- (0.5,0.5);
                            \draw[very thick, color=#1!50!white,fill=#1!50!white] (1.5,1.1) -- (1.9,1.4) -- (1.9,0.8) -- (1.5,1.1);
                            \draw[very thick, color=#1!50!white,fill=#1!50!white] (1.5,1.1) -- (1.1,1.4) -- (1.1,0.8) -- (1.5,1.1);
                            \draw[very thick, fill=#1!50!white,color=#1!60!white] (1.5,1.1) -- (1.6,0.5) -- (1.4,0.5) -- (1.5,1.1);

                            \draw[very thick, color=#1] (0.7, 2.1) -- (0.6, 2.4);
                            \draw[very thick, color=#1] (2.3, 2.1) -- (2.4, 2.4);
                            \draw[very thick, color=#1] (1.15, 2.2) -- (1.1, 2.5);
                            \draw[very thick, color=#1] (1.85, 2.2) -- (1.9, 2.5);

                        }
                }
        }
        \resizebox{\linewidth}{!}{
            \begin{tikzpicture}
                \node (syscenter) {};
                \node[rectangle, draw, below=0.3cm of syscenter] (env) {\scriptsize Env};
                \node[rectangle, draw, above=0.3cm of syscenter] (controller) {\scriptsize Contr};
                \node[rectangle, draw, left=0.3cm of syscenter] (act) {\scriptsize Actuator};
                \node[rectangle, draw, right=0.3cm of syscenter] (sensor) {\scriptsize Sensor};

                \node[rectangle, draw, dotted, very thick, fit=(env)(controller)(act)(sensor)] (Sys) {};
                \node[below=0.2cm of Sys] {\small System};
                \node[rectangle, draw, right=.8cm of Sys, minimum height=2.4cm, minimum width=.8cm] (monitor) {\rotatebox{90}{Monitor}};
                \node[rectangle, draw, minimum height=1.1cm, right=1cm of monitor, yshift=0.7cm, align=center, minimum width=1.4cm] (model) {HMM};
                \node[rectangle, draw, minimum height=1.1cm, right=1cm of monitor, yshift=-0.7cm, minimum width=1.4cm] (inference) {\scriptsize{inference}};

                \node[rectangle, below=0.2cm of monitor] (alarm) {\textcolor{green!50!black}{ok}/\textcolor{red!50!black}{alarm}};
                \draw[->] (env) edge[bend right=45] (sensor);
                \draw[->] (sensor) edge[bend right=45] (controller);
                \draw[->,dashed] (controller) edge[bend right=45] (sensor);

                \draw[->] (controller) edge[bend right=45] (act);
                \draw[->] (act) edge[bend right=45] (env);

                \draw[->] (sensor) -- node[above] {\scriptsize{obs}} (monitor);
                \draw[->] (monitor) -- (alarm);

                \draw[->] (monitor.317) -- node[above,align=center] {\scriptsize{obs}\\[-5pt] \scriptsize{trace}} (inference.west |- monitor.317);
                \draw[<-] (monitor.305) -- node[below,align=center] {\scriptsize{risk}\\[-5pt] \scriptsize{level}} (inference.west |- monitor.305);

                \draw[<-,dotted] (model) -- node[right] {\scriptsize{uses}} (inference);
                \draw[->,dotted] (model) -- +(0,0.7) -| node[pos=0.25,above] {\scriptsize{models}} (Sys);

            \end{tikzpicture}
        }
        \vspace{-6mm}
        \caption{White-box monitoring of systems}
        \label{fig:monitors}
    \end{minipage}\hfill
    \begin{minipage}{.47\textwidth}
        \resizebox{\linewidth}{!}{
            \begin{tikzpicture}

                \node[rectangle, minimum height=2.5cm, draw] (learner) {\rotatebox{90}{\scriptsize{learner}}};
                \node[rectangle,right=2cm of learner, yshift=0.7cm,align=center, draw, minimum height=1cm, minimum width=1.9cm] (mq) {\scriptsize{Membership?}\\[-3pt] \scriptsize{(the monitor)}};
                \node[rectangle,right=2cm of learner, yshift=-0.7cm,align=center, draw, minimum height=1cm, minimum width=1.9cm] (eq) {\scriptsize{Equivalence?}};
                \node[rectangle, right=0.8cm of eq, draw, minimum width=1.4cm, minimum height=1cm] (inc) {\scriptsize{Verifier}};
                \node[rectangle, right=0.8cm of mq, draw, minimum width=1.4cm, minimum height=1cm,align=center] (model) {HMM};
                \node[above=.2cm of model] (dummy) {};

                \draw[->] (learner.75) -- node[above,align=center] {\scriptsize{observation}\\[-5pt]\scriptsize{trace}} (mq.west |- learner.75);
                \draw[<-] (learner.72) -- node[below,align=center] {\scriptsize{ok/alarm}} (mq.west |- learner.72);
                \draw[->] (learner.285) -- node[above,align=center] {\scriptsize{state machine/}\\[-5pt]\scriptsize{set o.~traces}} (eq.west |- learner.285);
                \draw[<-] (learner.283) -- node[below,align=center] {\scriptsize{accept/reject}} (eq.west |- learner.283);
                \draw[->] (eq) --node[right] {\scriptsize sample} (mq);

                \draw[->] (eq) --node[above] {\scriptsize invoke} (inc);
                \draw[->, dotted] (mq) --node[above] {\scriptsize{uses}} (model);
                \draw[->, dotted] (inc) --node[right] {\scriptsize{uses}} (model);

            \end{tikzpicture}}
        \vspace{-2.5mm}
        \caption{Learning monitors.}
        \label{fig:learning}
    \end{minipage}
\end{figure}

\paragraph{Monitoring with HMMs.} Monitoring with HMMs assumes that a monitor receives a trace from the system and performs inference on the HMM modelling the system (see \cref{fig:monitors}). In the inference step, the key task is to estimate whether the current system state is dangerous, based on the available information in the form of a trace. Intuitively, the \emph{risk of a trace}~\cite{jungesRuntimeMonitorsMarkov2021} quantifies how likely it is that the system state is dangerous. Formally, this can be defined as the probability of ending in a dangerous state, conditioned on the fact that the system execution matches the trace. For a given trace, we may compute this risk, e.g., either via model checking~\cite{baierComputingConditionalProbabilities2014} or by a (forward) filtering that tracks a distribution over the current states~\cite{jungesRuntimeMonitorsMarkov2021,rabinerTutorialHiddenMarkov1989}. We call a trace \emph{unsafe} if its risk exceeds an acceptable threshold. Monitors should raise alarms only for unsafe traces.

\paragraph{What are correct monitors?}
Like in~\cite{acetoProbabilisticMonitorability2022}, we summarize the behavior of monitors by the    set of traces (i.e., a formal language) on which they raise an alarm. A monitor is correct iff it raises an alarm on all unsafe traces. We highlight that a monitor can be correct without doing inference at run time~\cite{cleavelandConservativeSafetyMonitors2023}! The key verification problem in this paper asks whether a monitor, represented as a deterministic finite automaton, accepts (i.e., raises an alarm on) exactly the unsafe traces. In this paper, we only consider traces that are bounded by some fixed horizon.

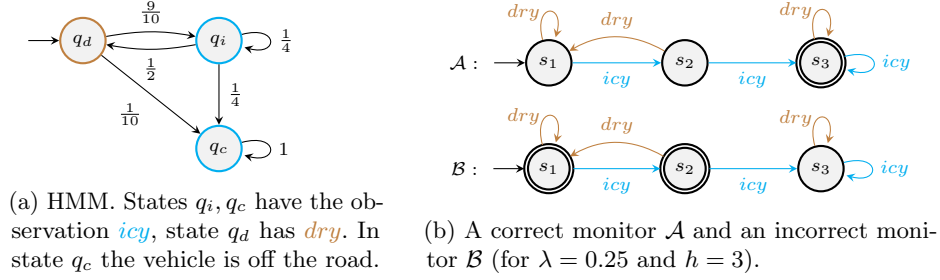
\begin{figure}[t]
    \begin{subfigure}{0.4\textwidth}
        \begin{tikzpicture}[node distance=1.8cm,every node/.style={font=\scriptsize}]
            \node[state,initial, dry] (qd) {$q_d$};
            \node[state, icy] (qi) [right of=qd] {$q_i$};
            \node[state, icy] (qc) [below=0.8cm of qi] {$q_c$};

            \draw[->] (qd) edge[bend left=10] node[above] {$\frac{9}{10}$} (qi);
            \draw[->] (qd) edge node[below left] {$\frac{1}{10}$} (qc);
            \draw[->] (qi) edge[bend left=10] node[below] {$\frac{1}{2}$} (qd);
            \draw[->] (qi) edge[loop right] node {$\frac{1}{4}$} ();
            \draw[->] (qi) edge node[right] {$\frac{1}{4}$} (qc);
            \draw[->] (qc) edge[loop right] node {$1$} ();
        \end{tikzpicture}
        \caption{HMM. States $q_i, q_c$ have the observation $\icy$, state $q_d$ has $\dry$. In state $q_c$ the vehicle is off the road.}
        \label{fig:centralhmm}
    \end{subfigure}
    \hfill
    \begin{subfigure}{0.55\textwidth}
        \center
        \begin{tikzpicture}[node distance=1.8cm,every node/.style={font=\scriptsize}]
            \node[state,initial] (q0) {$s_1$};
            \node (A) [left= 0.5cm of q0] {$\dfa:$};
            \node[state] (q1) [right of=q0] {$s_2$};
            \node[state, accepting] (q2) [right of=q1] {$s_3$};

            \draw[->] (q0) edge[loop above, dry] node [left] {$\dry$} ();
            \draw[->] (q0) edge[icy] node[below] {$\icy$} (q1);
            \draw[->] (q1) edge[dry, bend right] node[above] {$\dry$} (q0);
            \draw[->] (q1) edge[icy] node[below] {$\icy$} (q2);
            \draw[->] (q2) edge[icy, loop right] node {$\icy$} ();
            \draw[->] (q2) edge[dry, loop above] node[left] {$\dry$} ();

            \begin{scope}[yshift=-1.4cm]
                \node[state,initial,accepting] (q0) {$s_1$};
                \node (B) [left= 0.5cm of q0] {$\dfaB:$};
                \node[state, accepting] (q1) [right of=q0] {$s_2$};
                \node[state] (q2) [right of=q1] {$s_3$};

                \draw[->] (q0) edge[loop above, dry] node[left] {$\dry$} ();
                \draw[->] (q0) edge[icy] node[below] {$\icy$} (q1);
                \draw[->] (q1) edge[dry, bend right] node[above] {$\dry$} (q0);
                \draw[->] (q1) edge[icy] node[below] {$\icy$} (q2);
                \draw[->] (q2) edge[icy, loop right] node {$\icy$} ();
                \draw[->] (q2) edge[dry, loop above] node[left] {$\dry$} ();
            \end{scope}
        \end{tikzpicture}
        \caption{A correct monitor $\dfa$ and an incorrect monitor~$\dfaB$ (for $\lambda=0.25$ and $h=3$).
        }
        \label{fig:centralmon}
    \end{subfigure}
    \caption{Running example: HMM (a) and two monitors (b). }
    \label{fig:centralexample}
    \vspace{-.1cm}
\end{figure}

\paragraph{Illustrative toy example.}
As a running example, we consider an oversimplified car driving scenario, loosely inspired by runtime monitors obtained from high-fidelity simulations~\cite{DBLP:conf/atva/TorfahXJVS22}. A car can be in three states: It can be on a dry road, on an icy patch, or it has drifted off the road. The HMM in \cref{fig:centralhmm} describes how a car alternates between dry and icy road segments, and where being on an icy (dry) segment positively affects the probability that the next segment is icy (dry). When on an icy road, there is a higher probability to drift off the road. Our sensor can detect dry roads, but cannot distinguish between icy roads and off-road conditions. For a trace $\tau$ such as $\dry \cdot \icy \cdot \icy$, we can define the trace risk as the probability to be off-road conditioned on $\tau$, $\sfrac{13}{22}$ (for this trace, see \cref{ex:riskoftrace}).
In \cref{fig:centralmon}, monitor $\dfa$ is a finite automaton that describes the set of traces that end with two consecutive icy patches.
The verification question in this paper is now: \emph{Is this a correct monitor for threshold $0.25$?} That is, is an alarm raised iff the trace risk is above $0.25$?



\paragraph{Model-based approach.}
In this paper, we assume we are given a \emph{useful} model of the system.  While such white-box methods can be efficient and explainable, they require a model of the system, which is not always realistic. Alternatively, model-free approaches~\cite{cairoliNeuralPredictiveMonitoring2021} require no such assumptions, but they have weaker guarantees. Black-box methods~\cite{babaeeEmphPreventPredictiveRunTime2018} fit in between, they learn a model of the system and then use a white-box algorithm to provide the monitor. While we focus on the white-box approach, the results thus contribute towards black-box settings.

\paragraph{Computational complexity results.}
Deciding whether a monitor is correct is coNP-complete (\cref{lem:complexity}), assuming that the horizon is unary encoded.  In particular, we study the dual problem that asks whether a monitor misclassifies at least one trace as safe/unsafe.
A simpler variant of this problem asks whether there is a bounded trace in the HMM that is unsafe.
This problem is already strongly NP-hard (\cref{lem:ctpdecnphard}) and the optimization problem that asks to compute (the risk of) most risky trace is APX-hard (\cref{lem:ctpapxhard}), which indicates that it is intractable to even approximate this risk.

\paragraph{Our verification approach.}
While deciding whether a monitor is correct is in general not tractable, we suggest utilizing recent advances in the synthesis of probabilistic systems~\cite{andriushchenkoPAYNTToolInductive2021}.
Our presentation focusses on proving the absence of \emph{missed alarms}, i.e., we concentrate on showing that the monitor correctly identifies every unsafe trace, but we show that a similar reduction can be used to show that a monitor correctly identifies the safe traces.
First, for a single trace, computing the risk can be reduced to computing a reachability probability in an MC that is some kind of product between a deterministic finite automata (DFA) that accepts exactly the trace and the HMM~\cite{jungesRuntimeMonitorsMarkov2021}.
Thus, inspired by~\cite{DBLP:conf/tacas/BadingsVJSJ24}, we reduce our problem to the question: \emph{Is there a DFA (accepting exactly one trace, accepted by our monitor)
    such that the probability in the product MC exceeds a threshold?}
The answer is no iff the monitor has no missed alarms.
We formalize the problem using colored MDPs and solve it using (exact) probabilistic model synthesis methods~\cite{DBLP:conf/uai/Andriushchenko022}, as implemented in the tool PAYNT~\cite{andriushchenkoPAYNTToolInductive2021}.

\paragraph{Our learning approach.}
Ultimately, we do not only want to verify monitors, but we want to \emph{synthesize} correct monitors.
We already mentioned that the monitors are formal languages. When considering bounded traces, these are regular, and can thus be captured using DFAs. Thus, we aim to synthesize DFAs.
We do this using active automata learning~(AAL, \cite{DBLP:journals/iandc/Angluin87,DBLP:journals/cacm/Vaandrager17}), which is similar to other oracle-guided inductive synthesis loops~\cite{DBLP:journals/acta/JhaS17}.
Consider \cref{fig:learning}:
To use AAL, we must provide a \emph{membership oracle} that decides for a single trace whether the risk of the trace exceeds a threshold and an \emph{equivalence oracle} that decides whether a given hypothesis monitor is indeed correct. The membership oracle can be implemented via inference for standard monitors on HMMs (see above), while the equivalence oracle can be implemented by verifying the correctness of the monitor, as popularized in \emph{black-box checking}~\cite{DBLP:journals/jalc/PeledVY02}.
Practically, some minor modifications are necessary to embed our approach in an AAL framework: We have to handle the finite horizon, ignore traces that cannot occur, and combine our equivalence oracle with a conformance oracle to boost performance.

\paragraph{Contributions.}
In summary, this paper provides a first framework to learn runtime monitors---encoded as finite automata---that are verified to be correct with respect to a given HMM. The main contributions are:
\textbf{(1)}~We solve the verification problem by exhaustively searching for a counterexample using probabilistic system synthesis (\cref{sec:monver}).
\textbf{(2)}~We learn monitors by using the verifier above to answer equivalence queries in conjunction with active automata learning (\cref{sec:learning}).
\textbf{(3)}~We prove hardness of the verification problem (\cref{sec:compl}).
\textbf{(4)}~We demonstrate the feasibility and limitations of both the verifier and the learner on a series of benchmarks (\cref{sec:experiments}). Our prototype finds monitors that are provably correct on $10^{22}$ traces by verifying HMMs with thousands of states, up to a hundred observations, and monitors with over 100 states.

\smallskip
\noindent
The extended edition \appcite{\cite[App. A and B]{TR}}{\cref{app:proofs,app:complexity}} contains proof (sketches) for all lemmas and theorems.

\section{Formal Problem Statement}
Let $X$ be a finite set. A distribution $\mu$ over $X$ is a mapping $\mu\colon X \rightarrow [0,1]$ such that $\sum_{x \in X} \mu(x) = 1$. The set of all distributions over $X$ is written $\Distr(X)$.

\paragraph{DFAs.}
A \emph{deterministic finite automaton} (DFA) is a tuple $\dfa\defeq\left(\StsD, \alphabet, \trans, \init, \final\right)$. $\StsD$ is a finite set of \emph{states}, $\alphabet$ is an \emph{alphabet}, $\trans\colon \StsD \times \alphabet \rightharpoonup \StsD$ is a (partial) \emph{transition function}, $\init$ is the \emph{initial state}, and $\final \subseteq \StsD$ is a set of \emph{accepting} states\footnote{$\final$ is named after the synonymous final states to avoid confusion with actions.}.
Let $\varepsilon$ denote the empty word. We lift the transition relation to words, which is defined recursively as: $\trans^*(q, \varepsilon) \defeq \varepsilon$ if $a \in \alphabet$ and $\trans^*(q, w \cdot a) \defeq \trans(\trans^*(q,w), a)$. The \emph{language} $\lang{\dfa}$ of a DFA $\dfa$ is the set of all words that end in a final state of $\dfa$, $\lang{\dfa} \defeq \{ w \in \alphabet^* \mid \trans^*(\init, w) \in \final \}$. We say that $\dfa$ accepts $w$ if $w \in \lang{\dfa}$.

\subsection{Models}
We introduce MDPs and HMMs: The former are integral to our approach and the latter are crucial to the problem statement. Details can be found in \cite{baierPrinciplesModelChecking2008}.

\begin{definition}[MDP]
    A \emph{Markov decision process} (MDP) is a tuple $\left(\Sts, \init, \Act, \ptrans \right)$ with a countable nonempty set  $\Sts$ of \emph{states},  the \emph{initial state} $\init \in \Sts$, and the \emph{partial transition function} $\ptrans\colon \Sts \times \Act \rightharpoonup \Distr(\Sts)$.
\end{definition}
We use $\Act(s) \defeq \{ \act \mid \ptrans(s, \act) \neq \bot \}$ as the set of \emph{enabled actions}. We assume no deadlocks, i.e., for every $s \in \Sts$, $\Act(s) \neq \emptyset$.
A path $\pi$ is a (possibly infinite) sequence $s_0\cdot\act_0\cdots \in \Sts \times ((\Sts \times \Act)^{*} \cup (\Sts \times \Act)^{\omega})$, such that $a_i \in \Act(s_i)$ and $P(s_i, a_i)(s_{i+1}) > 0$ for every $i \geq 0$. The last state of a finite path is denoted by $\pi_\downarrow$.
The set of paths in MDP $\mdp$ is denoted as $\Pi^\mdp$, the set of finite paths is $\Pi_{fin}^\mdp$, the set of paths of at most length $h$ is $\Pi_h^\mdp$, and the set of paths of exactly length $h$ are $\Pi_{=h}^\mdp$. We write $\ptrans(s,a,s')$ for $\ptrans(s,a)(s')$.

A \emph{Markov chain} (MC) is an MDP where $|\Act(s)| = 1$ for every state $s \in \Sts$. We simplify notation and write MCs as a tuple $\left(\Sts, \init,  \ptrans \right)$, $\ptrans(s)$ to refer to the unique distribution $\ptrans(s, \act)$, and $\ptrans(s,s')$ for $\ptrans(s)(s')$. Paths in an MC are sequences of (only) states.
As in~\cite{baierPrinciplesModelChecking2008},
the probability measure $\Prob^{\mc}$ of an MC $\mc$ is the unique probability measure following from the canonical $\sigma$-algebra associated with $\mc$.
A reachability property on target states $T$ is the set of paths which contain a state $t \in T$.
The reachability probability $\Prob(\lozenge T)$ for $\lozenge T$ is defined using the standard cylinder set construction.
\begin{definition}[HMMs]
    A (risk-labelled) HMM is a tuple $\left(\Sts, \init, \ptrans, \Obs, \obsf, r \right)$ such that $\left(\Sts, \init, \ptrans \right)$ is an MC, $\Obs$ is a finite set of \emph{observations}, $\obsf\colon \Sts \rightarrow \Obs$ is the (deterministic) \emph{observation function}\footnote{We use deterministic observation functions for concise definitions. Stochastic observation functions can be expressed via a blow-up of the HMM, see e.g., \cite{jungesRuntimeMonitorsMarkov2021}.}, and $r\colon \Sts \to \RR_{\geq 0}$ is the \emph{risk function}.
\end{definition}
%
%
%
%
%
Notions such as paths are lifted from MCs. Furthermore, a \emph{trace} $\tau$ is a sequence of observations. We lift $\obsf$ from states to paths.
We define  $\Prob(\tau\mid\pi) \defeq 1$ if $\obsf(\pi) = \tau$ and zero otherwise. The probability of a trace $\tau$ is $\sum_{\pi\in\Pi^\mc} \Prob(\pi) \cdot \Prob(\tau\mid\pi)$.
Finally, the conditional probability on a trace $\tau\in\lang{\mc}$ is defined using Bayes' rule
$\Prob(\pi\mid\tau) \defeq \nicefrac{\Prob(\tau\mid\pi) \cdot \Prob(\pi)}{\Prob(\tau)}$.
We define $\lang{\mc} \defeq \{ \obsf(\pi) \mid \pi \in \Pi^\mc\}$.
\begin{example}
    We consider the HMM from \cref{fig:centralhmm} and $\tau = \dry\cdot \icy\cdot \icy$. The conditional probability $\Prob(q_d\cdot q_i\cdot q_i\ |~\tau)$ is $\sfrac{\Prob(\tau\ |~q_d\cdot q_i\cdot q_i)\cdot \Prob(q_d\cdot q_i\cdot q_i)}{\Prob(\tau)}$. $\Prob(\tau~|~q_d\cdot q_i\cdot q_i)$ is 1, and $\Prob(q_d\cdot q_i\cdot q_i)$ is $\sfrac{9}{40}$. The sum of the probabilities of all paths which observe $\tau$ is $\sfrac{11}{20}$. Thus, $\Prob(q_d\cdot q_i\cdot q_i\mid\tau)$ is $\sfrac{9}{22}$.
\end{example}

\subsection{Formal Problem Statement}
\begin{definition}[Monitor]
    A DFA $\dfa$ is a monitor for HMM $\mc$ if the alphabet for $\dfa$ coincides with the observations in $\mc$.
\end{definition}

\noindent Monitors should accept unsafe traces, which we define via their risk (level)~\cite{jungesRuntimeMonitorsMarkov2021}:
\begin{definition}[Trace risk, safe/unsafe traces]\label{def:safeunsafe}
    Given HMM $\mc$, the \emph{risk} of $\tau \in \lang{\mc}$ is: \[\textstyle R(\tau) \defeq \sum\nolimits_{\pi\in\Pi_{|\tau|}^\mc}\Prob(\pi \ |\ \tau) \cdot r(\pi_{\downarrow}).\]
    Let $\lambda_s \leq \lambda_u \in \mathbb{R}_{\geq 0}$ be the \emph{safe threshold}  and the \emph{unsafe threshold}, respectively.
    A trace $\tau \in \lang{\mc}$ with $R(\tau) > \lambda_u$ is \emph{unsafe}, while $\tau$ is \emph{safe} if $R(\tau) < \lambda_s$.
\end{definition}
We deliberately do not require $\lambda_s = \lambda_u$.
By picking $\lambda_s < \lambda_u$, some traces are neither safe nor unsafe, also called inconclusive.
We write $\safetraces{\mc,\lambda_s}{\leq h}$ (and $\unsafetraces{\mc,\lambda_u}{\leq h}$) for the set of safe (and unsafe) traces of length at most $h$.
\begin{example}\label{ex:riskoftrace}
    We consider the HMM from \cref{fig:centralhmm}, with the risk function assigning $1$ to $q_c$ and $0$ to all other states. Taking the trace $\dry \cdot \icy \cdot \icy$, there are three paths which could generate this trace. Two paths end in $q_c$, one ends in $q_i$. The paths ending in $q_c$ have a conditional probability of $\sfrac{13}{22}$. Since only these paths have a non-zero risk, the risk of the trace is $\sfrac{13}{22} \cdot r(q_c) = \sfrac{13}{22}$.
\end{example}
\begin{definition}[Missed/False alarms]
    Given a monitor $\dfa$ for HMM $\mc$, a horizon $h$, and thresholds $\lambda_s \leq \lambda_u \in \mathbb{R}_{\geq 0}$, the set of \emph{missed alarms} is $\missedalarms{\mc,\lambda_u}{\leq h}\!(\dfa) \defeq \unsafetraces{\mc,\lambda_u}{\leq h} \setminus \lang{\dfa}$. The set of \emph{false alarms} is $\falsealarms{\mc,\lambda_s}{\leq h}\!(\dfa) \defeq \safetraces{\mc,\lambda_s}{\leq h} \cap \lang{\dfa}$.
\end{definition}
\begin{definition}[Correct monitor]\label{def:cormon}
    Given thresholds $\lambda_s,\lambda_u$ and horizon $h$, a monitor $\dfa$ for HMM $\mc$ is \emph{correct} if $\missedalarms{\mc,\lambda_u}{\leq h}\!(\dfa) = \emptyset = \falsealarms{\mc,\lambda_s}{\leq h}\!(\dfa)$.
\end{definition}
A correct monitor raises an alarm for all unsafe traces and for no safe trace, i.e., missed alarms are false negatives, while false alarms are false positives.
\begin{corollary}
    A monitor $\dfa$ is correct iff $ \unsafetraces{\mc,\lambda_u}{\leq h} \subseteq \lang{\dfa} \subseteq \alphabet^* \setminus \safetraces{\mc, \lambda_s}{\leq h}$.
\end{corollary}
\begin{mdframed}\textbf{Problem statements.} 
    Given HMM $\mc$, thresholds $\lambda_s,\lambda_u$ and horizon $h$:
    \begin{enumerate}[topsep=2pt]
        \item\label{prob:MA} Given monitor $\dfa$ for $\mc$, are there  missed alarms, i.e., is $\missedalarms{\mc,\lambda_u}{\leq h}\!(\dfa)\! = \emptyset$?
        \item\label{prob:FA} Given monitor $\dfa$ for $\mc$, are there false alarms, i.e., is $\falsealarms{\mc,\lambda_s}{\leq h}\!(\dfa)\! = \emptyset$?
        \item\label{prob:Learn} Find a correct monitor $\dfa$ for $\mc$ w.r.t.\ $\lambda_s, \lambda_u$ and $h$.
    \end{enumerate}
\end{mdframed}
Problems 1 and 2 together allow checking whether a monitor is correct.
Furthermore, a correct monitor must exist, as $\unsafetraces{}{\leq h}$ is finite and thus regular.

\begin{example}
    We discuss monitor correctness for the example from \cref{fig:centralexample} using the correct monitor $\dfa$.
    Given the risk function assigning $1$ to $q_c$ and $0$ to all other states, the traces $\tau_1 = \dry \cdot \icy \cdot \icy$ and $\tau_2 = \dry \cdot \icy$ have risks $\sfrac{13}{22}, \sfrac{1}{10}$ respectively. If $\lambda_s$ is $\sfrac{1}{4}$ and the horizon $h$ is 3, $\tau_2$ is the trace with maximum risk not accepted by the monitor. Given that its risk is below $\lambda_s$, the monitor does not have any missed alarms. Similarly, monitor $\dfa$ does not have any false alarms for $\lambda_u = \sfrac{1}{4}$. Thus, $\dfa$ is a correct monitor for $\mc$ with $h$, $\lambda_s$, and $\lambda_u$.
\end{example}

\section{Monitor Verification}
\label{sec:monver}
We present our approach to the monitor correctness problem, which reduces checking the existence of missed alarms to the well-studied policy synthesis problem on colored MDPs, defined below. We first formalize this  policy synthesis problem and then present the step-wise transformation. Here, we focus on showing that there are no missed alarms  of exactly the length of the horizon (an adaption of \cref{prob:MA}). At the end of the section, we generalize our construction to traces of length \emph{at most} the horizon (\cref{prob:MA}) and to
finding false alarms (\cref{prob:FA}).

\subsection{Relating Missed Alarms to Color-Consistent Policies}
\label{sec:overview}\label{ssec:maintheorem}
A (memoryless) \emph{policy} for an MDP $\mc$ is a function $\sigma\colon S \to \Act$, which selects actions for every state.
An MDP and a policy induce an MC by only keeping the state-action pairs in the transition function selected by the policy.
We denote the probability measure on the MC induced by a policy $\sigma$ of the MDP $\mc$ as $\Prob^{\mc}_{\sigma}$
Policy synthesis for an MDP of a (reachability) property $\phi$ entails finding a policy for an MDP such that the induced MC entails $\phi$. Colored (aka: labelled) MDPs~\cite{DBLP:journals/jair/AndriushchenkoCMJK25} are an extension to MDPs that allow expressing dependencies between states that policies must adhere to. The following definition suffices for our needs:
\begin{definition}[Colored MDP]\label{def:colMDP}
    Given an MDP $\mdp$ with states $\Sts$, a \emph{colored MDP} is a tuple $\mdpc \defeq (\mdp, C, c)$, where $C$ is a set of \emph{colors}, and $c\colon \Sts \to C$. 
\end{definition}
\begin{definition}[Color consistent]
    \label{def:colconsistent}
    A memoryless policy $\sigma$ for a colored MDP $\mdpc$ is \emph{color consistent}\footnote{Colored MDPs with color-consistent policies coincide with memoryless policies for partially observable MDPs. However, POMDPs often consider history-dependent (belief-based) policies. We use \emph{colored MDPs} to avoid any confusion.} if for states $s, s'$,
    $c(s) = c(s')$ implies $\sigma(s) = \sigma(s')$.
\end{definition}
The set of all color-consistent policies is denoted $\ccSched$. Policy synthesis for colored MDPs asks to find a color-consistent policy such that the reachability probability to a set of target states is above a certain threshold, or to report that no such policy exists. This problem is NP-hard~\cite{DBLP:conf/aaai/ChatterjeeCD16}, but efficient heuristics exist~\cite{andriushchenkoPAYNTToolInductive2021}.
Policy synthesis for colored MDPs can prove the absence of missed alarms:
\begin{theorem}
    \label{thm:maintheorem}
    Given an HMM $\mc$, monitor $\dfa$, unsafe threshold $\lambda_u$, and horizon $h$, there is a colored MDP $\mdpc$ with target state $T$ and threshold $\lambda$ s.t.\
    \[ \exists \sigma \in \ccSched.\ \Prob^{\mdpc}_{\sigma}(\lozenge T) \geq \lambda\quad\text{ iff }\quad \exists \tau\in \missedalarms{\mc,\lambda_u}{= h}\!(\dfa).\]
\end{theorem}
Our proof, outlined in this section, is constructive and we show that we can use the construction to find a $\tau \in \missedalarms{}{}(\dfa)$, whenever such a $\tau$ exists.

\paragraph{Outline of the proof.}
The proof is a direct consequence of \cref{lem:toctp,lem:toactp,lem:tocoloredmdp} below. We observe that on the left-hand side of \cref{thm:maintheorem}, the monitor, horizon, observations, and risk do not occur, they must be encoded into the colored MDP.
Furthermore, while missed alarms are defined using conditional probabilities,  the policy synthesis problem is over reachability probabilities.
We describe our transformation in several steps. In \cref{ssec:ctp}, we encode the monitor into the HMM and transform the HMM to include both the horizon and the risk. In \cref{ssec:polsyn}, we resolve the conditioning and replace observations by nondeterminism.

\begin{corollary}
    There exists a map $t(\sigma) = \tau$, which, given a color-consistent policy $\sigma$, finds its associated trace $\tau$.
\end{corollary}

\subsection{The (Acyclic) Conditional Trace Risk Problem}\label{ssec:ctp}
\begin{figure}[t]
    \begin{minipage}[c]{.41\linewidth}
        \resizebox{\linewidth}{!}{
            \begin{tikzpicture}[node distance=.5cm and .5cm,every node/.style={font=\scriptsize}]
                \node[state,initial, dry] (qd1) {$\langle d,1\rangle$};
                \node[state, icy] (qi2) [right=of qd1] {$\langle i,2\rangle$};
                \node[state, icy] (qc2) [below=of qi2] {$\langle c,2\rangle$};
                \node[state, icy, accepting] (qi3) [right=of qi2] {$\langle i,3\rangle$};
                \node[state, icy, accepting] (qc3) [below=of qi3] {$\langle c,3\rangle$};
                \node[state, dry, accepting] (qd3) [right=of qi3] {$\langle d,3\rangle$};

                \draw[->] (qd1) edge[bend left] node[above] {$\frac{9}{10}$} (qi2);
                \draw[->] (qd1) edge node[below] {$\frac{1}{10}$} (qc2);

                \draw[->] (qi2) edge[bend left] node[above] {$\frac{1}{2}$} (qd1);
                \draw[->] (qi2) edge node[above] {$\frac{1}{4}$} (qi3);
                \draw[->] (qi2) edge node[above, near end] {$\frac{1}{4}$} (qc3);
                \draw[->] (qc2) edge node[above] {$1$} (qc3);

                \draw[->] (qc3) edge[loop right] node {$1$} ();
                \draw[->] (qi3) edge[loop above] node[above] {$\frac{1}{4}$} ();
                \draw[->] (qi3) edge node[right] {$\frac{1}{4}$} (qc3);
                \draw[->] (qi3) edge[bend right] node[above] {$\frac{1}{2}$} (qd3);
                \draw[->] (qd3) edge[bend right] node[above] {$\frac{9}{10}$} (qi3);
                \draw[->] (qd3) edge node[below, near start] {$\frac{1}{10}$} (qc3);
            \end{tikzpicture}}
        \caption{The HMM for \cref{ex:hmmproduct}. States are named by the HMM state, $\{d,i,c\}$ and the monitor state, $\{1,2,3\}$. The alarm states are marked accepting.}
        \label{fig:ctphmm}
    \end{minipage}\hfill
    \begin{minipage}[c]{.56\linewidth}
        \resizebox{\linewidth}{!}{
            \begin{tikzpicture}[node distance=.4cm and .4cm,every node/.style={font=\tiny},every loop/.style={looseness=2}]
                \node[state, other, initial, dry] (qd1) {$\left\langle 1, \atop \langle d,1\rangle \right\rangle$};

                \node[state, other, icy] (qi2) [right = of qd1] {$\left\langle 2, \atop \langle i,2\rangle \right\rangle$};


                \node[state, other, icy] (qc2) [below= of qi2] {$\left\langle 2, \atop \langle c,2\rangle \right\rangle$};

                \node[state, other, icy] (qi3) [right = of qi2] {$\left\langle 3, \atop \langle i,3\rangle \right\rangle$};

                \node[state, other, dry] (qd3) [above= of qi3] {$\left\langle 3, \atop \langle d,1\rangle \right\rangle$};

                \node[state, other, icy] (qc3) [below= of qi3] {$\left\langle 3, \atop \langle c,3\rangle \right\rangle$};

                \node[state, other]  (safe) [right =  of qi3] {$\safeas[4]$};
                \node[state, other]  (alarm) [below= of safe] {$\alarmas[4]$};
                \node[state, other, draw=gray] (ignore) [above= of safe] {$\igas[4]$};

                \draw[->] (qi3) edge node[above] {$1$} (safe);

                \draw[->] (qd3) edge node[above] {$1$} (ignore);

                \draw[->] (qc3) edge node[above] {$1$} (alarm);

                \draw[->] (qd1) edge node[above] {$\frac{9}{10}$} (qi2);
                \draw[->] (qd1) edge node[below] {$\frac{1}{10}$} (qc2);

                \draw[->] (qi2) edge node[above, near start] {$\frac{1}{2}$} (qd3);
                \draw[->] (qi2) edge node[above, near end] {$\frac{1}{4}$} (qi3);
                \draw[->] (qi2) edge node[below, near start] {$\frac{1}{4}$} (qc3);

                \draw[->] (qc2) edge node[below] {$1$} (qc3);

                \draw[->] (safe) edge[loop right] node[right] {$1$} ();
                \draw[->] (alarm) edge[loop right] node[right] {$1$} ();
                \draw[->] (ignore) edge[loop right] node[right] {$1$} ();
            \end{tikzpicture}}
        \caption{The HMM for \cref{ex:actp}, brown and cyan are \dry and \icy observations. Black is the new $\en$ observation, and gray is the new $\ig$ observation. All states are named with the step $i$, model state $s$ and monitor state $j$ as $\left\langle i, \langle s,j\rangle \right\rangle$.}
        \label{fig:actp}
    \end{minipage}
\end{figure}

First, we show how asking for a missed alarm can be rephrased into the conceptually simpler \emph{conditional trace risk} (CTR) problem.
We will further simplify the problem such that we are left with a problem on acyclic HMMs.


\subsubsection{CTR problem}
We build (a mild variation of) a standard product construction~\cite{baierPrinciplesModelChecking2008} between HMM and the DFA.  We define the \emph{alarm states} $T$ as those states which correspond to non-accepting states in the DFA. This is equivalent to taking a product with the complement of the monitor.
\begin{example}
    \label{ex:hmmproduct}
    \cref{fig:ctphmm} shows the product of the HMM and monitor $\dfaB$ from \cref{fig:centralexample}. Starting in the initial states $d$ and $1$, the HMM transitions to the $i$ state with probability $\sfrac{9}{10}$. This is an $\icy$ state, and thus the monitor takes the $\icy$ transition to state~$2$. In the product, this is a transition from $\langle d, 1\rangle$ to $\langle i, 2\rangle$ with probability $\sfrac{9}{10}$. The alarm states are any product states of the form $\langle \_ ,3 \rangle$.
\end{example}
\begin{definition}[HMM product]
    \label{def:product}
    Given an HMM $\mc=(\Sts, \iota^\mc, \ptrans, Z, \obsf, r)$ and monitor $\dfa=(Q, \Sigma, \delta, \iota^\dfa, F)$, the \emph{product HMM} $\langle\mc_{\times \dfa}, T\rangle$ is the HMM  $\mc_{\times \dfa} \defeq (S \times Q, \langle\iota^\mc,\iota^\dfa\rangle, \ptrans', Z, \obsf', r')$ with $\obsf'(\langle s, q\rangle) \defeq \obsf(s)$,  $r'(\langle s, q\rangle) \defeq r(s)$, $\ptrans'(\langle s, q\rangle, \langle s', \delta(q, \obsf(s'))\rangle) \defeq \ptrans(s, s')$ and $\ptrans'(x, x') \defeq 0$ otherwise, and finally the alarm states $T \defeq S\times F$.
\end{definition}
In the product, we can find a trace $\tau$ whose conditional trace risk exceeds a threshold iff $\tau$ is a missed alarm.
We state the decision problem that needs to be solved: It is key to our computational complexity analysis in \cref{sec:compl}.
\begin{definition}[CTR Decision Problem]
    \label{def:CTPdecprob}
    Given HMM $\mc$ with states~$S$ and risk~$r$, horizon $h$, alarm states $T\subseteq\Sts$, and threshold $\lambda_u \in \RR_{\geq 0}$, decide if \[\exists \tau \in \lang{\mc}.~ \sum\nolimits_{\pi\in\Pi^\mc_{=h}\mid\pi_\downarrow \in T}\Prob^\mc(\pi \mid \tau) \cdot r(\pi_\downarrow) > \lambda_u.\]
\end{definition}
We denote the set of witnesses $\tau$ to a CTR decision problem as $\mathsf{CTR}(\mc, h, T, \lambda_u)$.
The following lemma states the correctness of the transformation. It follows directly from the definition of missing alarms and the product with the complement.
\begin{restatable}{lemma}{toctp}
    \label{lem:toctp}
    Using the notation from \cref{thm:maintheorem}, \cref{def:product} and \cref{def:CTPdecprob}:
    \[ \exists \tau \in \missedalarms{\mc,\lambda_u}{= h}\!(\dfa) \quad\text{ iff }\quad \exists \tau \in \mathsf{CTR}(\mc_{\times\overline{\dfa}}, h, T, \lambda_u)\]
\end{restatable}

%

%

\subsubsection{ACTR problem}
We further simplify the problem by unrolling the model along the horizon $h$ and eliminating the risk function.
For the first $h$ steps, the unrolling is standard. At the horizon we transition to three dedicated states $\alarms$, $\safes$ and $\igs$\footnote{We add the $\igs$ state to easily modify the transformation for the no-false-alarms problem (See \cref{ssec:nfa}).} according to the risk and alarm states $T$.

\begin{example}
    \label{ex:actp}
    Consider the HMM $\mc$ depicted in \cref{fig:ctphmm}, which was defined in \cref{ex:hmmproduct}. We unroll $\mc$ with a horizon $h=3$. This yields the HMM in \cref{fig:actp}.
    The states are tuples of the unrolled step and the CTR state from $\mc$.
    The initial state becomes $\actps{1}{d}{1}$, which corresponds to a state at step $1$ and $\langle d,1 \rangle$ from the $\mc$. This state transitions to $\actps{2}{i}{2}$ and $\actps{2}{c}{2}$.
    States with a step value equal to the horizon are treated differently. From states at the horizon, we transition to either of three states that help classify the trace.
    Consider $\actps{3}{i}{3}$, it is at the horizon $h=3$, and it is in $T$.
    We want to classify such terminal states. As the (normalized) risk of $\actps{3}{i}{3}$ is $0$, the state is safe and $\actps{3}{i}{3}$ transitions with probability $1$ to $\safes[4]$. Next, Consider $\actps{3}{d}{1}$, $\langle d, 1\rangle$, it is not in $T$, but the state is at the horizon. This state can be ignored, and we transition to the ignore state, $\igs[4]$.
\end{example}

\begin{definition}[Unrolling with risk]
    \label{def:acyclichmm}
    The (risk-)unrolled HMM $\mc_{\blacktriangleright h}$ of an HMM $\mc = \left(\Sts, \init, \ptrans, \Obs, \obsf, r \right)$ with alarm states $T$ and horizon $h$ is the HMM \[ \mc_{\blacktriangleright h} \defeq (\{1,\hdots,h\} \times \Sts \cup \Sts_{\mathsf{end}}, \langle 1,\init\rangle, \ptrans', \Obs \cup \{\en, \ig\}, \obsf', r'),\] with $\Sts_{\mathsf{end}} \defeq  \{ \alarms, \safes, \igs \}$, $r'(\cdot) = 0$,
    $\obsf'$ given by (1) $\obsf'(\langle i, s\rangle) \defeq \obsf(s)$, \quad(2) $\obsf'(\igs) \defeq \ig$, \quad and (3) $\obsf'(\alarms) = \obsf'(\safes) \defeq \en$,
    and $\ptrans'$ given by:
    \begin{align*}
        \forall_{i\in\{1,\ldots,h-1\}}\quad \ptrans'(\langle i, s\rangle, \langle i+1, s'\rangle) & \defeq \ptrans(s,s'), \\
        \ptrans'(\langle h, s\rangle,\left\langle h+1, t\right\rangle)                            & \defeq
        \begin{cases}
            \frac{r(s)}{\max_{s\in\Sts} r(s)}     & \text{if } s \in F \text{ and } t = \talrm,  \\
            1 - \frac{r(s)}{\max_{s\in\Sts} r(s)} & \text{if } s \in F \text{ and } t = \tsafe,  \\
            1                                     & \text{if } s \notin F \text{ and } t = \tig, \\
            0                                     & \text{otherwise,}
        \end{cases}                      \\
        \ptrans'\left(\left\langle h+1, t\right\rangle, \left\langle h+1, t\right\rangle\right)   & \defeq 1.
    \end{align*}
\end{definition}

\begin{restatable}{lemma}{toactp}
    \label{lem:toactp}
    Given an HMM $\mc$, horizon $h$, alarm states $T$, and threshold $\lambda_u$, there exists a $\lambda\in(0,1]$ such that, using $\en$ and $\talrm$ from \cref{def:acyclichmm}:
    \[ \exists \tau \in \mathsf{CTR}(\mc, h, t, \lambda_u) \quad\text{ iff }\quad \exists \tau \in \lang{\mc_{\blacktriangleright h}}. \sum_{\pi\in\Pi^\mc}\Prob^\mc(\pi \cdot \talrm \mid \tau \cdot \en) \geq \lambda.\]
\end{restatable}
%
%

\subsection{Reduction to Consistent Policy Synthesis}\label{ssec:polsyn}
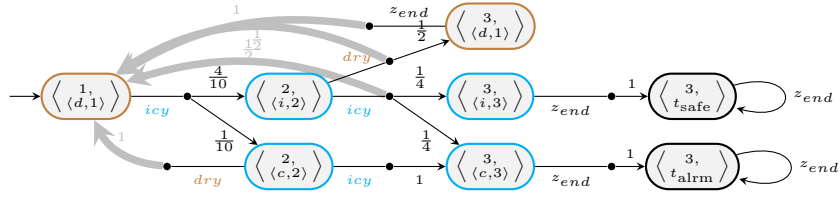
\begin{figure}[t]
    \centering
    \begin{tikzpicture}[node distance=.3cm and 1cm,every node/.style={font=\tiny}]
        \node[state, other,initial, dry] (qd1) {$\left\langle 1, \atop \langle d,1\rangle \right\rangle$};

        \node[state, other, icy] (qi2) [right = 1.5cm of qd1] {$\left\langle 2, \atop \langle i,2\rangle \right\rangle$};
        \node[state, other, icy] (qc2) [below= of qi2] {$\left\langle 2, \atop \langle c,2\rangle \right\rangle$};

        \node[state, other, icy] (qi3) [right = 1.5cm of qi2] {$\left\langle 3, \atop \langle i,3\rangle \right\rangle$};
        \node[state, other, dry] (qd3) [above= of qi3] {$\left\langle 3, \atop \langle d,1\rangle \right\rangle$};
        \node[state, other, icy] (qc3) [below= of qi3] {$\left\langle 3, \atop \langle c,3\rangle \right\rangle$};

        \node[state, other] (safe) [right = 1.5cm of qi3] {$\safeas[3]$};
        \node[state, other] (alarm) [below= of safe] {$\alarmas[3]$};

        \node[action] (qd1i) at ($(qd1)!0.5!(qi2)$)  {};
        \node[action] (qi2d) at ($(qi2)!0.5!(qd3)$) {};
        \node[action] (qi2i) at ($(qi2)!0.5!(qi3)$)  {};
        \node[action] (qc2d) at ($(qc2)!0.6!(qd1|-qc2)$) {};
        \node[action] (qc2i) at ($(qc2)!0.5!(qc3)$) {};

        \node[action] (qi3e) at ($(qi3)!0.6!(safe)$) {};
        \node[action] (qd3e) at ($(qd3)!0.6!(qi2|-qd3)$) {};
        \node[action] (qc3e) at ($(qc3)!0.6!(alarm)$) {};

        \draw (qi3) edge node[below] {$\en$} (qi3e);
        \draw[->] (qi3e) edge node[above] {$1$} (safe);

        \draw (qd3) edge node[above] {$\en$} (qd3e);
        \draw[return, ->] (qd3e) edge[bend right=20] node[above] {$1$} (qd1);

        \draw (qc3) edge node[below] {$\en$} (qc3e);
        \draw[->] (qc3e) edge node[above] {$1$} (alarm);

        \draw (qd1) edge node[below] {$\icy$} (qd1i);
        \draw[->] (qd1i) edge node[above] {$\frac{4}{10}$} (qi2);
        \draw[->] (qd1i) edge node[below] {$\frac{1}{10}$} (qc2);


        \draw (qi2) edge node[above] {$\dry$} (qi2d);
        \draw[->] (qi2d) edge node[above] {$\frac{1}{2}$} (qd3);
        \draw[->,return] (qi2d) edge[bend right] node[below] {$\frac{1}{2}$} (qd1);
        \draw (qi2) edge node[below] {$\icy$} (qi2i);
        \draw[->] (qi2i) edge node[above] {$\frac{1}{4}$} (qi3);
        \draw[->] (qi2i) edge node[below] {$\frac{1}{4}$} (qc3);
        \begin{scope}[on background layer]
            \draw[->,return] (qi2i) edge[bend right=20] node[above] {$\frac{1}{2}$} (qd1);
        \end{scope}

        \draw (qc2) edge node[below] {$\icy$} (qc2i);
        \draw[->] (qc2i) edge node[below] {$1$} (qc3);
        \draw (qc2) edge node[below] {$\dry$} (qc2d);
        \draw[->, return] (qc2d) edge[bend left] node[above] {$1$} (qd1);

        \draw (safe) edge[loop right] node {$\en$} (safe);
        \draw (alarm) edge[loop right] node {$\en$} (alarm);
    \end{tikzpicture}
    \caption{The MDP from \cref{ex:policysynthesis}. Unreachable states are omitted, as are any actions which return to the initial state from every state at the same step.}
    \label{fig:mdpex}
\end{figure}

For Lemma~\ref{lem:toactp}, we must find a trace such that a conditional reachability probability exceeds a threshold. We reformulate this into a policy synthesis problem (\cref{sec:overview}). The transformation combines two ideas: First, in every state, the policy can select the next observation, loosely inspired by~\cite{DBLP:conf/tacas/BadingsVJSJ24}. Second, we reformulate conditional reachability probabilities into  reachability probabilities, as in~\cite{jungesRuntimeMonitorsMarkov2021,baierComputingConditionalProbabilities2014}.
\begin{example}
    \label{ex:policysynthesis}
    We transform the HMM from \cref{fig:actp} to the colored MDP in \cref{fig:mdpex}. Consider $\actps{2}{i}{2}$ in the  HMM.
    The next observation is either $\dry$ or $\icy$, which corresponds to the two actions from $\actps{2}{i}{2}$ in the colored MDP.
    The $\dry$ action transitions to the ensuing states with the $\dry$ observation.
    The remaining probability of $\sfrac{1}{2}$ for the $\dry$ action is directed to the initial state.
    This is similarly applied to the $\icy$ action.
    State $\actps{2}{c}{2}$ does not reach any $\dry$ states, we still add a $\dry$ action redirecting to the initial state, as \emph{all} states with the same step value must have the same actions.
    State $\actps{3}{d}{1}$ in the HMM only has a transition to an $\ig$ state. We fully remove the $\ig$ state and the $\ig$ observation in the MDP. Finally, we make sure all states with step 3 have the same actions, which explains the $\en$ action from state $\actps{3}{d}{1}$.
\end{example}

\begin{definition}\label{def:polsyntrans}
    Given the HMM from \cref{def:acyclichmm} \[ \mc_{\blacktriangleright h} \defeq (\Sts' \defeq \{1,\hdots,h\} \times \Sts \cup \Sts_{\mathsf{end}}, \langle 1,\init\rangle, \ptrans, \Obs' \defeq \Obs \cup \{\en, \ig\}, \obsf, r),\] we define the colored MDP $\mdp_{\gtrdot h} \defeq \left(\left(\Sts'\setminus\{\igs\}, \init, \Act, \ptrans' \right), C, c\right)$ with $\Act \defeq \Obs'\setminus\{\ig\}$, $C \defeq \{1,\ldots,h\}$, coloring $c$ s.t.\  $c(\langle i,s \rangle) \defeq i$, and
    \begin{align*}
        \ptrans'(\langle i, s\rangle, z, \langle j, q\rangle) \defeq
        \begin{cases}
            \ptrans(\langle i, s\rangle, \langle j, q\rangle)                                                                             & \text{if } \obsf(\langle j, q\rangle) = z, \\
            \sum_{(\langle k, q'\rangle)\in S \mid \obsf(\langle k, q'\rangle) \neq z} \ptrans(\langle i, s\rangle, \langle k, q'\rangle) & \text{if } (\langle j, q\rangle) = \iota , \\
            0                                                                                                                             & \text{otherwise}
        \end{cases}
    \end{align*}
\end{definition}
Thus, we transition normally to a state if the action and observation of the target state are equal, otherwise we set the transition probability to zero. All the remaining probability mass is redirected towards the initial state\footnote{In the implementation, we can prune actions where, from every state with the same color, the action redirects to the initial state.}.

The above construction allows for conditioning on a trace $\tau$ by constructing a policy $\sigma$ that selects the $i$th observation of $\tau$ in the state with step $i$.
\begin{definition}[Trace consistent policy]
    Given a colored MDP $\mdp_{\gtrdot h}$ as in \cref{def:polsyntrans} and a trace $\tau \in (Z')^\star$, a \emph{trace consistent policy} satisfies $\sigma_\tau(\langle i, s \rangle) \defeq \tau^{(i)}$, where $\tau^{(i)}$ is the $i$th observation in $\tau$, for $i \leq |\tau|$ and $\sigma_\tau(\langle i, s \rangle) \defeq \en$ otherwise.
\end{definition}
Using the coloring as described in \cref{def:polsyntrans}, the trace consistent policies coincide with color-consistent policies (\cref{def:colconsistent}). Finding a missed-alarm trace now reduces to solving the color-consistent policy synthesis problem for targets $\{ \alarms \}$.

\begin{restatable}{lemma}{tocoloredmdp}
    \label{lem:tocoloredmdp}
    Given an HMM $\mc$, horizon $h$ alarm states $T$, and threshold $\lambda$, it holds that
    \vspace{-.1cm}
    \[
        \begin{array}{c}
            \exists \tau \in \lang{\mc_{\blacktriangleright h}}. \sum_{\pi\in\Pi^{\mc_{\blacktriangleright h}}}\Prob^{\mc_{\blacktriangleright h}}(\pi \cdot \alarms \mid \tau \cdot \en) \geq \lambda \\
            \Updownarrow                                                                                                                                                                               \\
            \exists \sigma \in \ccSched.  \Prob^{\mdp_{\gtrdot h}}_{\sigma}(\lozenge \{\alarms\}) \geq \lambda
        \end{array}
    \]
    Where $\sigma$ is trace consistent for $\tau$.
\end{restatable}

Note that the colored part of the colored MDPs is absolutely necessary in order to map a policy of the colored MDP to a trace.
\begin{example}
    Consider the MDP from \cref{fig:mdpex} while ignoring the coloring. A function mapping $\left\langle 2, \langle i,2\rangle \right\rangle$ to $\icy$ and $\left\langle 2, \langle c,2\rangle \right\rangle$ to $\dry$ is a valid policy for the MDP. However, there does not exist a trace which describes the actions taken in the model, i.e., the trace starts with $\icy$ but then sometime contains $\icy$ and sometimes $\dry$.
\end{example}
Restricting our policies to color-consistent policies makes sure in every step of the model we only ever take one action, corresponding with one observation in the trace at that step.

\subsection{Adapting to No-False-Alarms and Smaller Traces}\label{ssec:nfa}

\begin{figure}[t]
    \begin{subfigure}{.48\linewidth}
        \begin{tikzpicture}[node distance=.4cm and 1cm,every node/.style={font=\tiny}]
            \node[state, initial] (init) {$\iota$};

            \node[state, cloud, cloud puffs = 12, minimum width = 2cm, cloud puff arc=120] (Ml) [right = 1cm of init] {$\mdp_{\gtrdot l}$};
            \node[state, cloud, cloud puffs = 12, minimum width = 2cm, cloud puff arc=120] (Mh) [below = .5cm of Ml] {$\mdp_{\gtrdot h}$};

            \node (tvdots) [above=.1cm of Ml] {\vdots};
            \node at ($(Ml.south)!0.35!(Mh.north)$) {\vdots};

            \node[action] (Mla) at ($(init)!0.4!(Ml)$) {};
            \node[action] (Mha) at ($(init)!0.4!(Mh)$) {};

            \draw (init) edge node[above] {$l$} (Mla);
            \draw[->] (Mla) edge node[above left] {} (Ml);
            \draw (init) edge node[above] {$h$} (Mha);
            \draw[->] (Mha) edge node[above left] {} (Mh);
        \end{tikzpicture}
        \caption{A colored MDP combining the colored MDPs from \cref{def:polsyntrans} for horizons up to $h$, $\{\mdp_{\gtrdot l}\mid l \in \{1,\ldots,h\}\}$, into one MDP. }
        \label{fig:leqstepa}
    \end{subfigure}\hfill
    \begin{subfigure}{.48\linewidth}
        \begin{tikzpicture}[node distance=.4cm and 1.2cm,every node/.style={font=\tiny}]
            \node[state, initial] (init) {$\iota$};

            \node[state] (i1) [right= of init] {$\langle \iota, 1\rangle$};
            \node[state] (il) [right= .5cm of i1] {$\langle \iota, l\rangle$};
            \node[state] (ih) [right= .5cm of il] {$\langle \iota, h\rangle$};
            \node at ($(i1)!0.5!(il)$) {\dots};
            \node at ($(il)!0.5!(ih)$) {\dots};

            \node[state, cloud, minimum width = 3.5cm, cloud puffs = 18, cloud puff arc=120] (Mh) [below =of il] {$\mdp_{\gtrdot h}$};

            \draw[->] (i1) -- (Mh.puff 3);
            \draw[->] (Mh.puff 2) -- (il);
            \draw[->] (il) -- (Mh.puff 18);
            \draw[->] (Mh.puff 18) -- (ih);
            \draw[->] (ih) -- (Mh.puff 16);

            \node[action] (i1a) at ($(init)!0.5!(i1)$) {};
            \node[action, yshift=.5cm] (ila) at (i1a) {};
            \node[action, yshift=1cm] (iha) at (i1a) {};

            \draw (init) --node[below, near end] {$1$} (i1a);
            \draw[->] (i1a) -- (i1);

            \draw (init) --node[below, near end] {$l$} (ila);
            \draw[->] (ila) edge[bend left] (il);

            \draw (init) --node[above, near end] {$h$} (iha);
            \draw[->] (iha) edge[bend left] (ih);
        \end{tikzpicture}
        \caption{A bisimilar colored MDP of \cref{fig:leqstepa}, containing only $\mdp_{\gtrdot h}$. States $\langle \iota, 1\rangle$ through $\langle \iota, h\rangle$ from $\mdp_{\gtrdot h}$ are shown separate in order to show their relation to $\iota$.}
        \label{fig:leqstepb}
    \end{subfigure}
    \caption{Transformation steps needed for \cref{thm:leqhtheorem}.}
    \label{fig:leqhorizon}
    \vspace{-.3cm}
\end{figure}
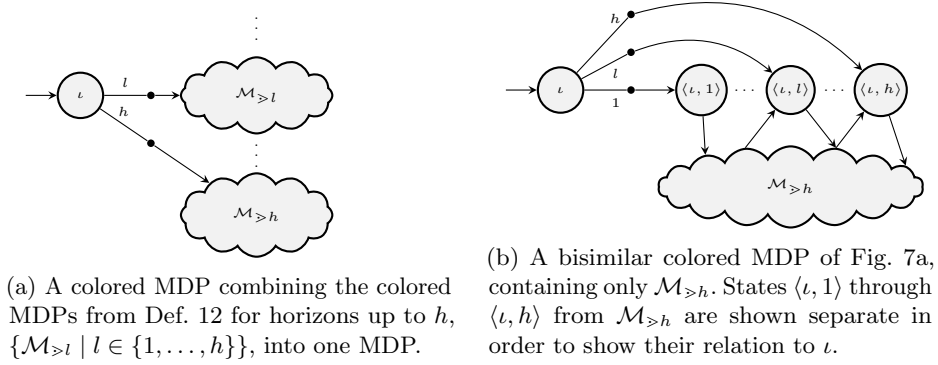

\paragraph{Traces of length at most the horizon.}
The approach for \cref{thm:maintheorem} only works for traces of length exactly equal to the horizon. We generalize this approach to traces of length at most the horizon.
\begin{restatable}{theorem}{leqhtheorem}
    \label{thm:leqhtheorem}
    Given an HMM $\mc$, a monitor $\dfa$, unsafe threshold $\lambda_u$, horizon $h$, and risk $r$, there is a colored MDP $\mdpc$ with target states $T$, and threshold $\lambda$ s.t.\
    \[ \exists \sigma \in \ccSched.\ \Prob^{\mdpc}_{\sigma}(\lozenge T) \geq \lambda\quad\text{ iff }\quad \exists \tau\in \missedalarms{\mc,\lambda_u}{\leq h}\!(\dfa).\]
\end{restatable}
\noindent The main insight for this theorem is shown in \cref{fig:leqhorizon}. We combine the colored MDPs given by \cref{thm:maintheorem} for horizons 1 to $h$ into one colored MDP, such that a policy starts by choosing which length trace to use (\cref{fig:leqstepa}). We can also instead directly construct a bisimulation quotient  $\mc_{\gtrdot h}$ of this combined MDP  with a small addition (\cref{fig:leqstepb}). We detail this construction in \appcite{\cite[App. A]{TR}}{\cref{app:proofs}}.

\paragraph{Finding false alarms (solving \cref{prob:FA}).}
We modify the transformation from \cref{thm:leqhtheorem} such that it solves the no-false-alarms problem.
This problem differs in two ways from the no-missed-alarms problem. We are finding a trace \emph{accepted by} the monitor, and we find a trace whose risk is \emph{below} the \emph{safe threshold}.
\begin{restatable}{theorem}{nfa}
    \label{thm:nfa}
    Given an HMM $\mc$, a monitor $\dfa$, safe threshold $\lambda_s$, horizon $h$, and risk $r$, there is a colored MDP $\mdpc$ with target states $T$, and threshold $\lambda$ s.t.\
    \[ \exists \sigma \in \ccSched.\ \Prob^{\mdpc}_{\sigma}(\lozenge T) \geq \lambda\quad\text{ iff }\quad \exists \tau\in \falsealarms{\mc,\lambda_s}{\leq h}\!(\dfa).\]
\end{restatable}
\noindent We highlight the ideas here, for details see \appcite{\cite[App. A]{TR}}{\cref{app:proofs}}.
In order to find a trace accepted by the monitor, we no longer take the complement of the monitor while transforming to CTR. To find a safe trace we compute reachability on $\safes$ instead of $\alarms$ while taking as a threshold $1-\lambda$. Thus, we find a trace whose probability of being safe is above a threshold\footnote{We cannot directly aim to compute a trace whose risk is below a threshold since minimizing reachability of $\alarms$ will result in a scheduler that always resets to the initial state if possible. Consider the colored MDP in \cref{fig:mdpex}, the color-consistent policy with the lowest probability of reaching the alarm state is as follows: take the $\dry$ action in step 2, and $\en$ in step 3. This policy has probability zero of reaching either $\alarms$ or $\safes$, and thus does not map to a valid trace.}.

\section{Learning Correct Monitors}\label{sec:learning}
We present \emph{learning} correct monitors (\cref{prob:Learn}), by combining automata learning using a \emph{Minimally Adequate Teacher} (MAT,~\cite{DBLP:journals/iandc/Angluin87}) and monitor verification.

\paragraph{MAT framework.}
We briefly recap the MAT framework, for details see~\cite{DBLP:journals/cacm/Vaandrager17}.
A minimally adequate teacher answers two types of questions:
A \emph{membership query} (MQ), which in our setup means \emph{should a trace be accepted by the monitor?}, and an \emph{equivalence query} (EQ), \emph{is this monitor correct?}
Furthermore, if the answer to an equivalence query is negative, we must provide a counterexample that witnesses why the monitor is not correct.
Various algorithms implementing the MAT framework for DFA learning exist. For the purpose of this paper, we use the L$^{\star}$ algorithm \cite{DBLP:journals/iandc/Angluin87} to learn a monitor. The learner asks MQs to the teacher until a \emph{hypothesis monitor} can be constructed which is consistent with the MQs. Once such a hypothesis is constructed, its correctness is verified using an EQ.

\paragraph{Verification as a MAT.}
To learn a monitor $\dfa$, we provide the HMM $\mc$, a risk function $r$, a horizon $h$, a learning threshold $\lambda_l$, and the safe and unsafe thresholds $\lambda_s$ and $\lambda_u$. The  additional learning threshold $\lambda_l$ is used to define an MQ whenever $\lambda_s \neq \lambda_u$: In particular, for MQs, each trace must be unambiguously safe or unsafe: the MAT framework does not allow for flagging certain traces as \emph{don't care}, while  traces with a risk between $\lambda_s$ and $\lambda_u$ can be considered don't care in our setting.  Likewise, the MQ must also be defined for traces $\tau \not \in \lang{\mc}$ or traces longer than the horizon. We thus adapt the notion of safe traces from \cref{def:safeunsafe}.
\begin{definition}\label{def:toverMQ}
    Given any trace $\tau \in \Obs^\star$ and a horizon $h$, membership query  $\MQ_{\lambda_l}$ is a function such that $\MQ_{\lambda_l}(\tau)$ is unsafe iff $\tau \in \lang{\mc}$\footnote{Defining traces $\tau\notin\lang{\mc}$ as safe is an arbitrary design decision.} and $\tau \in \unsafetraces{\lambda_l}{\leq h}$.
\end{definition}
Such a function for $\MQ_{\lambda_l}$ can be defined by keeping track of the probability of being in each state after every observation from the trace \emph{or} by model checking the induced Markov chain that reflects the trace-consistent policy in \cref{sec:monver}~\cite{rabinerTutorialHiddenMarkov1989,jungesRuntimeMonitorsMarkov2021}.
For EQs, we simply use the notion of correctness from \cref{def:cormon}.
\begin{definition}\label{def:toverEQ}
    Given an HMM $\mc$, and a monitor $\dfa$, an $\EQ_{\lambda_s, \lambda_u}$ is a function $\EQ_{\lambda_s, \lambda_u}(\dfa)\in\{\top\}\cup\Obs^\star$, such that, $\EQ_{\lambda_s, \lambda_u}^\mc(\dfa)$ holds if $\dfa$ is correct for $\mc$ with $\lambda_s$, and $\lambda_u$ (in the sense of \cref{def:cormon}), and $\EQ_{\lambda_s, \lambda_u}^\mc(\dfa)$ returns a missed alarm or false alarm trace for an incorrect $\dfa$.
\end{definition}
The EQ requires checking both for no-missing-alarms, and for no-false-alarms. We perform these checks using our novel verification algorithm described in \cref{sec:monver}.

\begin{restatable}{lemma}{learncor}
    Given a MAT with a $\EQ_{\lambda_s, \lambda_u}$ and a $\MQ_{\lambda_l}$, a monitor learned with L$^\star$ is correct as long as $\lambda_s \leq \lambda_l \leq \lambda_u$.
\end{restatable}
\noindent When $\lambda_s < \lambda_u$, the EQ has an inconclusive area given by the interval $(\lambda_s, \lambda_u)$.
This means that our EQ does not check for equivalence, but simply accepts any correct monitor~\cite{DBLP:journals/jalc/PeledVY02}. We investigate the effect of this inconclusive area in \cref{sec:experiments}.

\paragraph{Conformance queries.}
An alternative to the EQ in \cref{def:toverEQ} is a conformance query \cite{groceAdaptiveModelChecking2002}. It tests a monitor by sampling traces from the HMM and checking if the MQ and the monitor agree. If the monitor and the MQ don't agree on a trace, it is given as a counterexample. In our approach we use a hybrid of conformance and verification EQs. Monitors produced early in the learning process often contain many missed alarms and false alarms. Verification can find them, however, applying the transformation from \cref{sec:monver} has a constant cost. Conformance queries can often find a counterexample faster if they have a high probability of occurring. We thus first try to find easy counterexamples using conformance, when this does not produce any we perform our verification routine.

\section{Computational Complexity}\label{sec:compl}
This section discusses the hardness of monitor verification (\cref{lem:complexity}) and the inapproximability of a related optimization problem (\cref{lem:ctpapxhard}).
\begin{theorem}
    \label{lem:complexity}
    \emph{Is a monitor correct?} (w.\ unary coded horizon) is coNP-complete.
\end{theorem}
In fact, we study the dual to this problem, i.e., checking the existence of a counterexample. We call this problem \emph{monitor co-verification}. For monitor co-verification, \emph{membership} in NP follows from false alarms or missed alarms (of length up to horizon) being the witnesses.
Verifying whether a trace is a false or missed alarm can be done in polynomial time, by checking whether the automaton accepts it and computing the trace risk (see \cref{sec:learning}).

To establish NP-hardness, we consider the CTR problem from~\cref{def:CTPdecprob}. As a solution to the monitor co-verification problem solves the CTR problem (using a trivial monitor), this implies NP-hardness of the former problem.
\begin{lemma}
    \label{lem:ctpdecnphard}
    The CTR Decision Problem is (strongly) NP-hard.
\end{lemma}
The proof features a reduction from CNF-SAT, the problem of satisfiability of a propositional formula. We illustrate the reduction, details are in \appcite{\cite[App. B]{TR}}{\cref{app:complexity}}.

\tikzset{
    >=stealth, 
    every state/.style={thick, fill=gray!10, minimum size=0.8cm,inner sep=1pt}, 
    initial text=$ $, 
    action/.style={circle, fill, inner sep=1pt},
    icy/.style={draw=cyan},
    dry/.style={draw=brown},
    return/.style={color=lightgray, line width=1mm},
    other/.style={rounded rectangle},
    problem/.style={rectangle, draw, minimum height=1cm, minimum width=1.5cm, align=center}
}

\begin{figure}[t]
    \begin{subfigure}{0.13\textwidth}
        \begin{tikzpicture}
            \node[circle, inner sep=4pt, initial, initial above, initial text=,draw] (s1) {};
            \node[right=0.3cm of s1] (ci) {$\vdots$};
            \node[rectangle, draw, above=0.8cm of ci] (c1) {$\mathcal{G}_{c_1}$};
            \node[rectangle, draw, below=0.8cm of ci] (cn) {$\mathcal{G}_{c_m}$};
            \draw[->] (s1) edge node[right] {$\nicefrac{1}{m}$} (c1);
            \draw[->] (s1) edge node[right] {$\nicefrac{1}{m}$} (cn);
        \end{tikzpicture}
        \caption{HMM.}
        \label{fig:gadgetsconnected}
    \end{subfigure}
    \begin{subfigure}{0.86\textwidth}
        \begin{tikzpicture}[every node/.style={font=\scriptsize}]

            \node[state,initial text=,initial] (s1) {$s_{1,j}$};
            \node[state,right=0.4cm of s1,yshift=0.5cm,draw=green] (s1y) {$s_{1,j}^\top$};
            \node[state,right=0.4cm of s1,yshift=-0.5cm,draw=red] (s1n) {$s_{1,j}^\bot$};
            \node[state,above=1.1cm of s1,] (s1p) {$s'_{1,j}$};
            \node[state,right=0.4cm of s1p,yshift=0.5cm,draw=green] (s1yp) {$s_{1,j}'^\top$};
            \node[state,right=0.4cm of s1p,yshift=-0.5cm,draw=red] (s1np) {$s_{1,j}'^\bot$};
            \node[state,right=1.8cm of s1,] (s2) {$s_{2,j}$};
            \node[state,right=0.4cm of s2,yshift=0.5cm,draw=green] (s2y) {$s_{2,j}^\top$};
            \node[state,right=0.4cm of s2,yshift=-0.5cm,draw=red] (s2n) {$s_{2,j}$};
            \node[state,above=1.1cm of s2,] (s2p) {$s'_{2,j}$};
            \node[state,right=0.4cm of s2p,yshift=0.5cm,draw=green] (s2yp) {$s_{2,j}'^\top$};
            \node[state,right=0.4cm of s2p,yshift=-0.5cm,draw=red] (s2np) {$s_{2,j}'^\bot$};
            \node[state,right=1.8cm of s2,] (s3) {$s_{3,j}$};
            \node[state,right=0.4cm of s3,yshift=0.5cm,draw=green] (s3y) {$s_{3,j}^\top$};
            \node[state,right=0.4cm of s3,yshift=-0.5cm,draw=red] (s3n) {$s_{3,j}^\bot$};
            \node[state,above=1.1cm of s3,] (s3p) {$s'_{3,j}$};
            \node[state,right=0.4cm of s3p,yshift=0.5cm,draw=green] (s3yp) {$s_{3,j}'^\top$};
            \node[state,right=0.4cm of s3p,yshift=-0.5cm,draw=red] (s3np) {$s_{3,j}'^\bot$};
            \node[state,right=1.8cm of s3,] (s4) {$s_{4,j}$};
            \node[state,above=1.1cm of s4,] (s4p) {$s'_{4,j}$};

            \node[state,right=0.4cm of s4,] (f) {$f$};
            \node[state,accepting, above=1.1cm of f,] (t) {$t$};
            \draw[->] (s4) edge (f);
            \draw[->] (s4p) edge (t);

            \draw[->] (s1) edge node[above] {$0.5$} (s1y);
            \draw[->] (s1) edge node[above] {$0.5$} (s1n);
            \draw[->] (s2) edge node[above] {$0.5$} (s2y);
            \draw[->] (s2) edge node[above] {$0.5$} (s2n);
            \draw[->] (s3) edge node[above] {$0.5$} (s3y);
            \draw[->] (s3) edge node[above] {$0.5$} (s3n);
            \draw[->] (s1p) edge node[above] {$0.5$} (s1yp);
            \draw[->] (s1p) edge node[above] {$0.5$} (s1np);
            \draw[->] (s2p) edge node[above] {$0.5$} (s2yp);
            \draw[->] (s2p) edge node[above] {$0.5$} (s2np);
            \draw[->] (s3p) edge node[above] {$0.5$} (s3yp);
            \draw[->] (s3p) edge node[above] {$0.5$} (s3np);

            \draw[->] (s1y) edge node[above] {$1$} (s2p);
            \draw[->] (s1n) edge node[above] {$1$} (s2);
            \draw[->] (s2y) edge node[above] {$1$} (s3);
            \draw[->] (s2n) edge node[above] {$1$} (s3);
            \draw[->] (s3y) edge node[above] {$1$} (s4);
            \draw[->] (s3n) edge node[above] {$1$} (s4p);
            \draw[->] (s1yp) edge node[above] {$1$} (s2p);
            \draw[->] (s1np) edge node[above] {$1$} (s2p);
            \draw[->] (s2yp) edge node[above] {$1$} (s3p);
            \draw[->] (s2np) edge node[above] {$1$} (s3p);
            \draw[->] (s3yp) edge node[above] {$1$} (s4p);
            \draw[->] (s3np) edge node[above] {$1$} (s4p);

        \end{tikzpicture}
        \caption{Gadget $\mathcal{G}_c$ demonstrated for $c = x_1 \lor\neg x_3$}
        \label{fig:hmmgadget}
    \end{subfigure}
    \caption{Illustrations for Lemma~\ref{lem:ctpdecnphard}, with $m$ clauses and variables $x_1, x_2, x_3$.}
\end{figure}
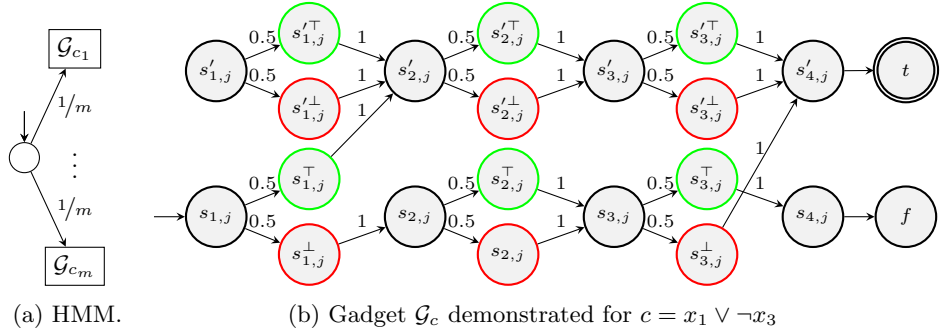

We construct HMM $\mc_\varphi$ from CNF $\varphi$ over variables $X$ such that there is a trace with risk $1$ iff there is a satisfying assignment to $\varphi$. In particular, that trace exists iff there is a trace $\tau$ s.t.\  all corresponding paths $\pi$ with $\obsf(\pi) = \tau$ reach state $t$.
The traces are of the form $\#\# \cdot \{ \bot, \top \} \cdot \# \cdot \{ \bot, \top \} \dots \cdot \#$: Trace $\#\# \cdot \alpha(x_0) \cdot \# \cdot \alpha(x_1) \dots \cdot \#$ represents an assignment  $\alpha\colon X \rightarrow \{\bot,\top \}$.
We construct $\mc_\varphi$ such that any trace that ensures reaching $t$ with probability one encodes a satisfying assignment for $\varphi$.  We create gadgets for every clause. The gadgets are connected as in \cref{fig:gadgetsconnected}:
That is, to ensure reaching $t$ along every path, we must reach $t$ in every gadget.
The gadget $\mathcal{G}_j$ intuitively `evaluates' $c_j$ with respect to an assignment, as exemplified in Fig.~\ref{fig:hmmgadget}.
A path (or its trace) through $\mathcal{G}_j$ `reads' variable $x_i$ in state $s_{i,j}$ and transitions to $s_{i,j}^\top$ or to $s_{i,j}^\bot$. The states are labelled $\#, \top, \bot$, respectively. However, for a trace where $\alpha(x_i) = \top$, only the former path corresponds to the trace (and symmetrically for $\alpha(x_i) = \bot$). That path reaches state $t$ iff the assignment satisfies at least one literal in the clause.

%

%

We now show that the construction above suffices to show that it is hard to approximate the maximal risk that a monitor admits.
\begin{definition}[CTR Optimization Problem]
    Given an HMM $\mc$ with states in $S$, a unary encoded horizon $h$, a set of alarm states $T\subseteq\Sts$: \[\max\nolimits_{\tau \in \lang{\mc}}~ \sum\nolimits_{\pi\in\Pi^\mc_{=h}\mid\pi_\downarrow \in T}\Prob^\mc(\pi \mid \tau) \cdot r(\pi_\downarrow).\]\end{definition}
\begin{lemma}
    \label{lem:ctpapxhard}
    The CTR Optimization Problem is APX-hard.
\end{lemma}
This follows from a strict reduction from MAX-3SAT, which is an inapproximable and APX-hard problem~\cite{DBLP:journals/jacm/Hastad01}. The construction coincides with the reduction in \cref{lem:ctpdecnphard} by observing that the conditional probability to reach a $t$ state is given by $\sfrac{1}{m}$ times the number of satisfied clauses, i.e., we can compute the maximal number of satisfied clauses in $\varphi$ using the HMM $\mc_\varphi$.

\section{Experiments}\label{sec:experiments}
\noindent We empirically evaluate the monitor verification (\cref{sec:monver}) and monitor learning (\cref{sec:learning}) algorithms using our prototype implementation called \alg.
Code, benchmarks, and logs will be publicly available via the artifact evaluation.

\paragraph{Setup.}
The \alg tool is implemented in Python and C++ on top of the model checker \textsc{Storm}~\cite{henselProbabilisticModelChecker2022} for data structures and for the MQs in \cref{sec:learning}~\cite{jungesRuntimeMonitorsMarkov2021}.
We use PAYNT~\cite{andriushchenkoPAYNTToolInductive2021} to verify colored MDPs (\cref{def:colMDP}), using exact arithmetic to avoid numerical problems on these types of benchmarks~\cite{DBLP:conf/tacas/HartmannsJQW23}.
The learner uses the AAlpy framework~\cite{muskardinAALpyActiveAutomata2022}.
All experiments are run  on a single thread of an AMD Ryzen TRP 5965WX, with a memory limit of 15 GiB, and a time-out of 12 hours.

\paragraph{Benchmarks.}
We take the benchmarks \textsc{Airport}, \textsc{Refuel}, \textsc{Evade}, and \textsc{Hidden-Incentive} from \cite{jungesRuntimeMonitorsMarkov2021}, adapted to HMMs. We add \textsc{Icy-Driving}, a scaled-up version of the running example and \textsc{SnL} based on the game ``Snakes and Ladders''.
While the benchmarks from the literature contain many observations, i.e., few states share an observation, the new benchmarks only have a few different observations. All benchmarks are scalable. The risk function is defined by a temporal property, e.g., the probability of reaching a bad state within a few steps.

\subsubsection{Efficiency of Monitor Verification}
We first investigate scalability along different dimensions and identify the bottlenecks of our verification method.

\paragraph{Setup.}
We verify the HMMs with respect to three monitors obtained during the learning experiments (below, with $\lambda_s = \lambda_u = 0.3$). Every version of the benchmark is run on the first (incorrect) monitor that passed a limited conformance check, an (incorrect) monitor obtained halfway through the learning process, and the final correct monitor.
We verify correctness w.r.t.\ the same $\lambda_s, \lambda_u$.




\begin{table}[t]
    \begin{adjustbox}{max width=\linewidth}
        \begin{scriptsize}
            \begin{tabular}{{@{}llrrrrrrrrrrrrr@{}}}
                \toprule
                                       &                & \multicolumn{7}{c}{Benchmark} & \multicolumn{6}{c}{\alg}                                                                                                                                                                                       \\
                \cmidrule(lr){3-10}\cmidrule(lr){11-15}
                                       &                & $h$                           & MA/FA                    & $|\Sts^\mc|$ & $|\ptrans^\mc|$ & $|Z|$ & $|\Sts^\dfa|$ & $|\ptrans^\dfa|$ & $|\mathcal{L}^{\leq h}|$ & Time (s)  & Trans (s) & PAYNT (s) & $|\mdp_{\gtrdot h}|$ & $\lambda^{found}$ \\
                \midrule
                \textsc{AirportB-7}    & \textsc{A-180} & 9                             & MA                       & 470          & 2550            & 50    & 66            & 3300             & $10^{12}$                & 5         & 1         & 4         & 2275                 & 0.69              \\
                \textsc{AirportB-7}    & \textsc{A-182} & 9                             & FA                       & 470          & 2550            & 50    & 66            & 3300             & $10^{12}$                & 3         & $\leq 1s$ & 3         & 2275                 & 0.10              \\
                \textsc{AirportB-7}    & \textsc{A-184} & 9                             & MA                       & 470          & 2550            & 50    & 319           & 15950            & $10^{12}$                & 23        & 1         & 22        & 3503                 & 0.41              \\
                \textsc{AirportB-7}    & \textsc{A-186} & 9                             & FA                       & 470          & 2550            & 50    & 319           & 15950            & $10^{12}$                & 17        & 1         & 15        & 3503                 & 0.24              \\
                \textsc{AirportB-7}    & \textsc{A-188} & 9                             & MA                       & 470          & 2550            & 50    & 569           & 28450            & $10^{12}$                & 367       & 2         & 365       & 5111                 & \checkmark        \\
                \textsc{AirportB-7}    & \textsc{A-190} & 9                             & FA                       & 470          & 2550            & 50    & 569           & 28450            & $10^{12}$                & 340       & 3         & 338       & 5111                 & \checkmark        \\
                \midrule
                \textsc{Evade}         & \textsc{E-40}  & 9                             & MA                       & 385          & 1473            & 325   & 197           & 64025            & $10^{14}$                & 34        & 26        & 4         & 680                  & 1.00              \\
                \textsc{Evade}         & \textsc{E-42}  & 9                             & FA                       & 385          & 1473            & 325   & 197           & 64025            & $10^{14}$                & 25        & 23        & 2         & 680                  & \checkmark        \\
                \midrule
                \textsc{Hidden-Incen.} & \textsc{H-2}   & 10                            & FA                       & 397          & 1649            & 100   & 106           & 10600            & $10^{18}$                & 10        & 3         & 6         & 1261                 & 0.30              \\
                \textsc{Hidden-Incen.} & \textsc{H-10}  & 10                            & FA                       & 397          & 1649            & 100   & 298           & 29800            & $10^{18}$                & 2972      & 4         & 2968      & 1411                 & \checkmark        \\
                \midrule
                \textsc{Icy-Driving}   & \textsc{I-34}  & 3                             & FA                       & 3            & 6               & 2     & 8             & 16               & $10^{0}$                 & $\leq 1s$ & $\leq 1s$ & $\leq 1s$ & 8                    & \checkmark        \\
                \textsc{Icy-Driving}   & \textsc{I-10}  & 10                            & FA                       & 3            & 6               & 2     & 2             & 4                & $10^{4}$                 & $\leq 1s$ & $\leq 1s$ & $\leq 1s$ & 29                   & \checkmark        \\
                \textsc{Icy-Driving}   & \textsc{I-22}  & 25                            & FA                       & 3            & 6               & 2     & 2             & 4                & $10^{11}$                & 1126      & $\leq 1s$ & 1126      & 74                   & \checkmark        \\
                \midrule
                \textsc{RefuelB}       & \textsc{R-20}  & 10                            & MA                       & 263          & 7127            & 72    & 31            & 2232             & $10^{22}$                & 3         & 1         & 2         & 1812                 & \checkmark        \\
                \textsc{RefuelB}       & \textsc{R-22}  & 10                            & FA                       & 263          & 7127            & 72    & 31            & 2232             & $10^{21}$                & 5         & 2         & 3         & 1812                 & \checkmark        \\
                \midrule
                \textsc{SnL}           & \textsc{S-76}  & 16                            & MA                       & 101          & 502             & 4     & 427           & 1708             & $10^{11}$                & 393       & $\leq 1s$ & 392       & 13547                & 0.54              \\
                \textsc{SnL}           & \textsc{S-78}  & 16                            & FA                       & 101          & 502             & 4     & 427           & 1708             & $10^{10}$                & 1418      & $\leq 1s$ & 1417      & 13547                & 0.00              \\
                \textsc{SnL}           & \textsc{S-80}  & 16                            & MA                       & 101          & 502             & 4     & 588           & 2352             & $10^{11}$                & 4601      & 2         & 4600      & 14381                & \checkmark        \\
                \textsc{SnL}           & \textsc{S-82}  & 16                            & FA                       & 101          & 502             & 4     & 588           & 2352             & $10^{10}$                & 2414      & 1         & 2413      & 14381                & \checkmark        \\
                \bottomrule
            \end{tabular}
        \end{scriptsize}
    \end{adjustbox}
    \vspace{.2cm}
    \caption{Subset of verification results found in \appcite{\cite[App. C.2]{TR}}{\cref{app:verrestable}}. The columns give the family name, an ID, horizon, and whether we check for missed alarms or false alarms. We give the size of the HMM (states, transitions), the number of observations, the size of the DFA (states, transitions), and the size of the language after pruning unreachable states. Furthermore, we list the run time for the complete verification procedure as well as the time spent on transforming the problem into a policy synthesis problem and the policy synthesis in PAYNT. Lastly, we list the size of the colored MDP produced by the transformation and the risk of the found counterexample. If no trace was found with a risk above (or below, for FA) the indicated threshold, the monitor is correct (\checkmark).}
    \label{tab:verifysubest}
    \vspace{-.5cm}
\end{table}

\paragraph{Results.}
We present our results in \cref{tab:verifysubest}, which is a subset of the 516 benchmarks shown in \appcite{\cite[App. C.2]{TR}}{\cref{app:verrestable}}. Generally, we observe that we verify the correctness of monitors on at least billions of traces, which shows that enumerating the traces is not a feasible alternative. Our verification handles monitors and HMMs with both hundreds of states and up to hundred thousands of transitions, see benchmarks \textsc{E-40}, \textsc{E-42}, \textsc{H-10}.
Benchmarks \textsc{A-180}, \textsc{A-182} reflect verification w.r.t. almost trivial monitors, for which it is typically easy to find a counterexample, \textsc{A-184}, \textsc{A-186}, \textsc{S-76}, \textsc{S-78} reflect a semi-correct monitor, and \textsc{A-188}, \textsc{A-190}, \textsc{S-80}, \textsc{S-82} reflect verification of the same HMM with respect to a larger correct monitor.
Increasing the horizon significantly increases the runtime, even for small models, e.g., \textsc{I-34} compared to \textsc{I-10} and \textsc{I-22}.
In all benchmarks, the runtime consists almost exclusively of creating the input to PAYNT (taking the product and creating the MDP) and in running PAYNT. The former runs in polynomial time in the size of the input (see \appcite{\cite[App. C.1]{TR}}{\cref{app:transtimeres}}), whereas the latter uses various heuristics to avoid the exponential computation time. In the current implementation, except for the (comparably) large \textsc{Evade} benchmarks, the vast majority of time is spent on PAYNT. The transformation is never the bottleneck (see \appcite{\cite[App. C.2]{TR}}{\cref{app:verrestable}}).


\subsubsection{Efficiency of Monitor Learning}
A key contribution of this paper is the ability to use verification for the EQs in monitor learning.
We consider the necessity of these EQs, the size of the learned monitors, and the efficiency of learning them, both for $\lambda_s = \lambda_u$ and $\lambda_s \neq \lambda_u$.

\paragraph{Setup.}
We use the MAT framework from \cref{sec:learning}.
Before every EQ, we run conformance checking (max.\ 100 samples using as threshold $\lambda_s$ and 100 samples with $\lambda_u$, see \cref{sec:learning}).
As hyper-parameters, we investigate (1) $\lambda_s = 0.1, \lambda_l = 0.3, \lambda_u = 0.35$, with inconclusive trace, and (2) $\lambda_s = \lambda_l = \lambda_u = 0.3$, without inconclusive traces%
\footnote{We study correctness of monitors learned by the baseline w.r.t.\ different $\lambda_l$ in \appcite{\cite[App. D]{TR}}{\cref{sec:ressampling}}.}.
We compare against a baseline that does not use EQs, i.e., the baseline uses the MAT framework with only conformance checking (max.\ 100000 samples, different numbers of samples are tested in \appcite{\cite[App. D]{TR}}{\cref{sec:ressampling}}).

\begin{figure}[t]
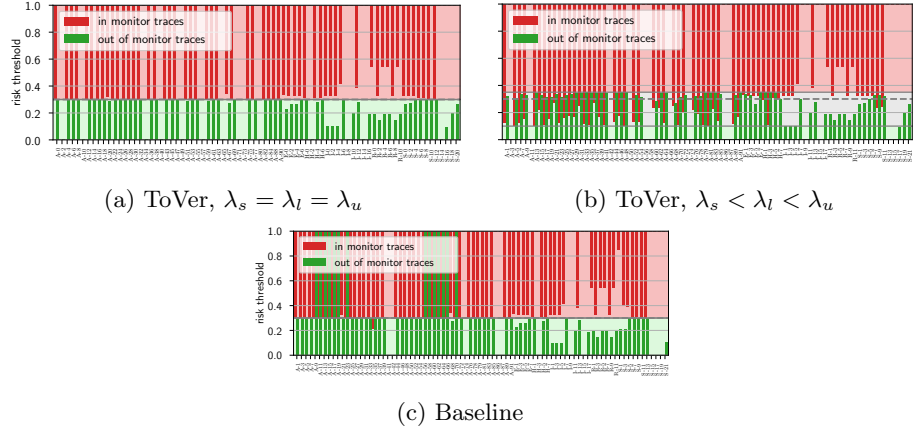

    \begin{subfigure}{.51\linewidth}
        \begin{adjustbox}{max width=\linewidth}
            \input{images/plots/thresholds_tover_no_slack.pgf}
        \end{adjustbox}
        \caption{\alg, $\lambda_s=\lambda_l=\lambda_u$}
        \label{fig:learnThrToverNS}
    \end{subfigure}\hfill
    \begin{subfigure}{.47\linewidth}
        \begin{adjustbox}{max width=\linewidth}
            \input{images/plots/thresholds_tover_slack.pgf}
        \end{adjustbox}
        \caption{\alg, $\lambda_s < \lambda_l < \lambda_u$}
        \label{fig:learnThrToverS}
    \end{subfigure}\hfill
    \centering
    \begin{subfigure}{.47\linewidth}
        \begin{adjustbox}{max width=\linewidth}
            \input{images/plots/theresholds_sampling.pgf}
        \end{adjustbox}
        \caption{Baseline}
        \label{fig:learnThrSampling}
    \end{subfigure}
    \caption{Actual alarm and actual no-alarm thresholds from monitors learned with \alg and baseline. The line between the green/gray is $\lambda_u$, the line between red/gray area is $\lambda_s$. The dotted line is $\lambda_l$. Missing bars reflect either time-outs or out-of-memory.}
\end{figure}

\paragraph{Are the monitors correct?}
Using verification in the EQ, we always learn correct monitors. We validate this experimentally \emph{and} show that the baseline does not always yield correct monitors. For every monitor we determine the unsafe trace with the lowest risk (\emph{actual alarm threshold}, $\lambda_u^{\min}$) and the safe trace with the highest risk (\emph{actual no-alarm threshold}, $\lambda_s^{\max}$). In a correct monitor, we have $\lambda_u^{\min} \geq \lambda_u$ and $\lambda_s^{\max} \leq \lambda_s$.
\cref{fig:learnThrToverS,fig:learnThrToverNS,fig:learnThrSampling} show $\lambda_u^{\min}$ and $\lambda_s^{\max}$ for \alg and for the baseline. Visually, a monitor is correct if its red bar never touches the green area and the green bar never touches the red area. In 14 out of 87 benchmarks the baseline learns a monitor that misses alarms. One baseline learned monitor has false alarms.

\begin{figure}[t]
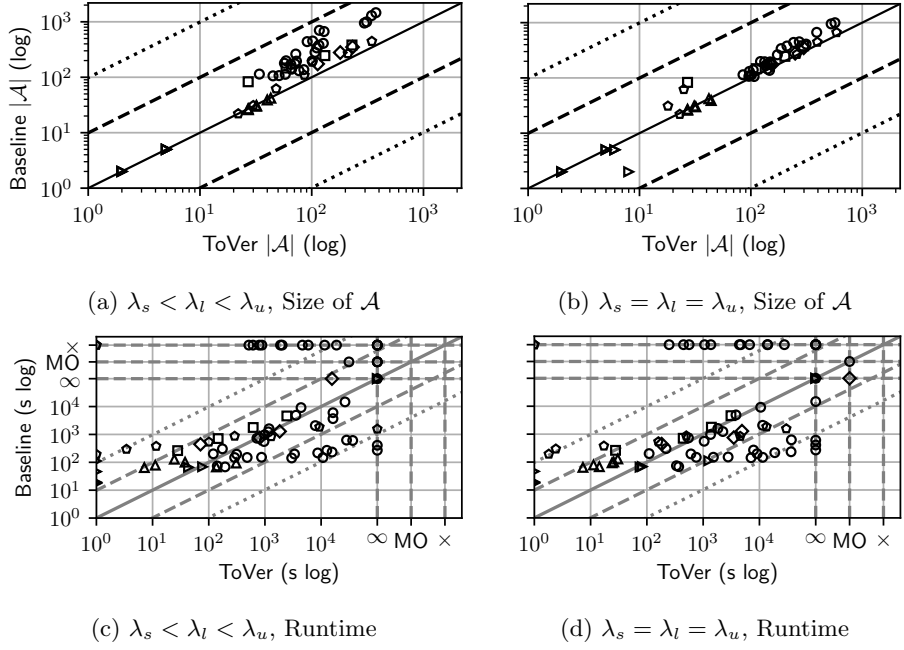

    \begin{subfigure}{.525\linewidth}
        \begin{adjustbox}{max width=\linewidth}
            \input{images/plots/monitor_sizes_with_slack.pgf}
        \end{adjustbox}
        \caption{$\lambda_s < \lambda_l < \lambda_u$, Size of $\dfa$}
        \label{fig:monsizes}
    \end{subfigure}\hfill
    \begin{subfigure}{.455\linewidth}
        \begin{adjustbox}{max width=\linewidth}
            \input{images/plots/monitor_sizes_no_slack.pgf}
        \end{adjustbox}
        \caption{$\lambda_s=\lambda_l=\lambda_u$, Size of $\dfa$}
        \label{fig:monsizens}
    \end{subfigure}\hfill
    \begin{subfigure}{.525\linewidth}
        \resizebox{\linewidth}{!}{
            \input{images/plots/runtimes_with_slack.pgf}}
        \caption{$\lambda_s < \lambda_l < \lambda_u$, \mbox{Runtime}}
        \label{fig:runtime-ws-comp}
    \end{subfigure}\hfill
    \begin{subfigure}{.455\linewidth}
        \resizebox{\linewidth}{!}{
            \input{images/plots/runtimes_no_slack.pgf}}
        \caption{$\lambda_s=\lambda_l=\lambda_u$, Runtime}
        \label{fig:runtime-ns-comp}
    \end{subfigure}
    \caption{States in learned monitors and runtimes: \alg vs. baseline. The runtime includes labels for time-outs ($\infty$), out-of-memory (MO), and incorrect ($\times$).}
\end{figure}

\paragraph{How big are the monitors?}
\alg{} learns monitors with hundreds of states and tens of thousands transitions, see \cref{fig:monsizes,fig:monsizens} (log-scale!) and \appcite{\cite[App. E.2]{TR}}{\cref{app:learnrestable}}. For the literature on AAL, these are large automata~\cite{yangImprovingModelInference2019,aichernigBenchmarkingCombinationsLearning2024,smeenkApplyingAutomataLearning2015}. Comparing the sizes of the monitors learned using \alg and the baseline, monitors are smaller (up to 8 times, mostly at least 1.5 times smaller)\footnote{For $\lambda_s = \lambda_u$, the language of correct monitors learned with \alg and baseline are equivalent up to the horizon, but the monitors respond differently on longer traces.}.

\paragraph{How fast do we learn the monitors?}
\begin{figure}[t]
    \resizebox{\linewidth}{!}{
        \input{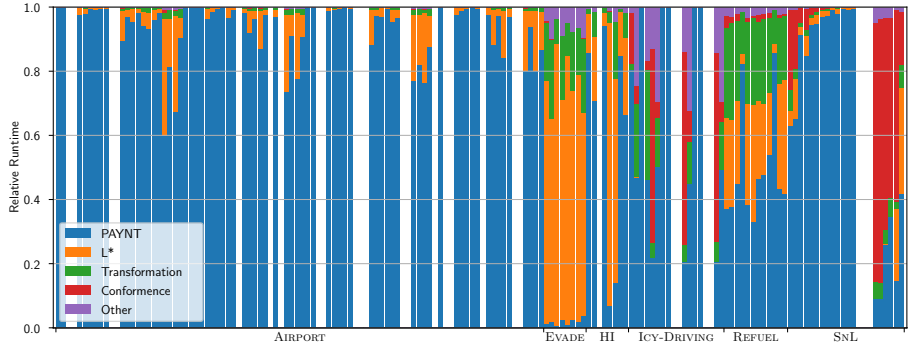}}
    \caption{Division of runtime for \alg learning for different models.
        \emph{PAYNT} is the time spent by PAYNT verifying colored MDPs. \emph{L$^\star$} is the time spent by the L$^\star$ algorithm creating a hypothesis including doing MQs. \emph{Transformation} is the time spent transforming HMMs and monitors into colored MDPs for PAYNT. \emph{Conformance} is the time spent on conformance testing during the EQs. Lastly, \emph{Other} denotes any time not spend in the aforementioned processes.
        Empty bars denote HMMs for which learning did not finish.}
    \label{fig:runtimediv}
\end{figure}
We compare the runtime of \alg and baseline in the \cref{fig:runtime-ws-comp,fig:runtime-ns-comp} (log-scale!).
We remark that only \alg{} is guaranteed to be correct. Neither of the two learning algorithms is clearly faster than the other, but \alg{} has the potential to significantly accelerate the learning process, despite the high complexity. One reason could be that \alg needs half or fewer EQs to learn a monitor as can be seen in \appcite{\cite[App. E.2]{TR}}{\cref{app:learnrestable}}. In \cref{fig:runtimediv}, we detail where the time is spent. For most benchmarks, the EQ (in particular, PAYNT) is the bottleneck. However, for several \textsc{Evade} benchmarks, most time is spent within L$^\star$ code. We conjecture this happens as finding counterexamples is simple in these models.


%


\paragraph{The role of an inconclusive area.}
We compare between $\lambda_s<\lambda_l<\lambda_u$ and $\lambda_s = \lambda_l = \lambda_u$, i.e., with and without an inconclusive area. The baseline does not actively support such an inconclusive area. With an inconclusive area, more monitors are correct (i.e., strictly speaking, we do not test equivalence but acceptance). The learner indeed finds monitors that are two to five times smaller (also compare \cref{fig:monsizes,fig:monsizens}). For benchmarks \textsc{Icy-Driving}, \textsc{Evade}, and \textsc{Airport}, this also translates to faster runtimes than using conformance checking, sometimes by orders of magnitudes.



\section{Related Work}
This work studies \emph{monitoring based on stochastic systems} and combines \emph{active learning} with \emph{probabilistic verification}. We consider related work in those directions.

\paragraph{Model-based monitoring for stochastic systems.}
Runtime verification is a wide field, see~\cite{DBLP:journals/sttt/FalconeFM12,sanchezSurveyChallengesRuntime2019,DBLP:series/lncs/HavelundRR19} for surveys. We review work on \emph{model-based} runtime monitoring for stochastic systems. In particular, using state estimation on HMMs to  decide whether to raise an alarm given a particular trace has been investigated in~\cite{sistlaMonitoringTemporalProperties2008,stollerRuntimeVerificationState2012,wilcoxRuntimeVerificationStochastic2010}, extended to hybrid models~\cite{sistlaRuntimeMonitoringStochastic2012}, models with nondeterminism~\cite{jungesRuntimeMonitorsMarkov2021} and randomly timed models~\cite{DBLP:conf/tacas/BadingsVJSJ24}. We use these techniques to answer membership queries. The HMMs for runtime monitoring can be learned from a set of traces, see, e.g.,~\cite{babaeeEmphPreventPredictiveRunTime2018,babaeeAcceleratedLearningPredictive2019} and more recently \cite{cleavelandConservativeSafetyMonitors2023}, where they find the state risks at design time using model checking and use state estimation for runtime verification. 
Related to runtime monitoring is runtime enforcement, in particular shielding~\cite{DBLP:conf/tacas/BloemKKW15,RW87,DBLP:conf/aaai/FultonP18,jansen-et-al-concur-2020}. Shielding is successful in fully observable models but less studied in partial observable settings, in~\cite{DBLP:conf/aaai/Carr0JT23}, shields are computed for qualitative properties.
Finally, in \cite{acetoProbabilisticMonitorability2022}, a more general notion of correct monitors via linear time $\mu$-calculus is investigated, while in \cite{DBLP:journals/corr/abs-2412-11754} a notion of correct \emph{predictors} is introduced.
Both can be seen as generalizations of our definitions.

\paragraph{Learning monitors.}
Learning monitors has been advocated in, e.g., \cite{cairoliNeuralPredictiveMonitoring2021,DBLP:conf/nfm/PhanG0PSS20,DBLP:conf/atva/TorfahXJVS22,zolfagharianSMARLASafetyMonitoring2025}. Closest to our setting is recent work in \cite{jungesActiveLearningRuntime2025}, which also uses state estimation for membership queries, but combines this with conformance queries and learns decision trees rather than automata. Crucially, by using conformance queries, the guarantees are significantly weaker, see also our experiments. 


\paragraph{Probabilistic verification.}
The verification of our monitors applies model checking of conditional probabilities~\cite{DBLP:conf/tacas/AndresR08,baierComputingConditionalProbabilities2014} to runtime verification, similar to \cite{jungesRuntimeMonitorsMarkov2021,DBLP:conf/tacas/BadingsVJSJ24}. Most related is recent work in~\cite{DBLP:conf/tacas/BadingsVJSJ24}, where the models are CTMCs and the observation trace is uncertain itself. They also encounter a notion of trace-consistent policies, but instead of using synthesis, they overapproximate the verification by considering all policies. In contrast, our method is \emph{complete}. Verification with partial observability as in our HMMs also occurs in the verification of partially observable MDPs~\cite{DBLP:journals/corr/abs-2405-13583}, which can also be tackled using synthesis approaches~\cite{DBLP:conf/cav/AndriushchenkoBCJKM23}. Finally, considering MDPs as \emph{distribution transformers} yields related but semantically different computationally hard problems that have been solved using (different) inductive synthesis approaches~\cite{DBLP:conf/cav/AkshayCMZ23,DBLP:conf/ijcai/0001CMZ24}.

\section{Conclusion and Future Work}
This paper presented a first approach to verification of monitors with respect to hidden Markov models. It embeds this verification procedure in an automata learning framework. The empirical evaluation is encouraging but also shows the limitations of the off-the-shelf frameworks. We see three avenues for future work: (1)~Dedicated synthesis methods for conditional probabilities and the specific structure of our colored MDPs. (2)~Automata learning for acyclic models and don't-care results.
(3)~Faster verification for longer (or even unbounded) traces.

\clearpage
\bibliographystyle{splncs04}
\bibliography{literature,zotero_dblp}

\begin{thebibliography}{10}
\providecommand{\url}[1]{\texttt{#1}}
\providecommand{\urlprefix}{URL }
\providecommand{\doi}[1]{https://doi.org/#1}

\bibitem{acetoProbabilisticMonitorability2022}
Aceto, L., Achilleos, A., Anastasiadi, E., Francalanza, A., Ing{\'{o}}lfsd{\'{o}}ttir, A., Lehtinen, K., Pedersen, M.R.: On probabilistic monitorability. In: Principles of Systems Design. Lecture Notes in Computer Science, vol. 13660, pp. 325--342. Springer (2022)

\bibitem{aichernigBenchmarkingCombinationsLearning2024}
Aichernig, B.K., Tappler, M., Wallner, F.: Benchmarking combinations of learning and testing algorithms for automata learning. Formal Aspects Comput.  \textbf{36}(1),  3:1--3:37 (2024)

\bibitem{DBLP:conf/cav/AkshayCMZ23}
Akshay, S., Chatterjee, K., Meggendorfer, T., Zikelic, D.: Mdps as distribution transformers: Affine invariant synthesis for safety objectives. In: {CAV} {(3)}. Lecture Notes in Computer Science, vol. 13966, pp. 86--112. Springer (2023)

\bibitem{DBLP:conf/ijcai/0001CMZ24}
Akshay, S., Chatterjee, K., Meggendorfer, T., Zikelic, D.: Certified policy verification and synthesis for {MDP}s under distributional reach-avoidance properties. In: {IJCAI}. pp. 3--12. ijcai.org (2024)

\bibitem{DBLP:conf/tacas/AndresR08}
Andr{\'{e}}s, M.E., van Rossum, P.: Conditional probabilities over probabilistic and nondeterministic systems. In: {TACAS}. Lecture Notes in Computer Science, vol.~4963, pp. 157--172. Springer (2008)

\bibitem{DBLP:journals/corr/abs-2405-13583}
Andriushchenko, R., Bork, A., Budde, C.E., Ceska, M., Grover, K., Hahn, E.M., Hartmanns, A., Israelsen, B., Jansen, N., Jeppson, J., Junges, S., K{\"{o}}hl, M.A., K{\"{o}}nighofer, B., Kret{\'i}nsk{\'{y}}, J., Meggendorfer, T., Parker, D., Pranger, S., Quatmann, T., Ruijters, E., Taylor, L., Volk, M., Weininger, M., Zhang, Z.: Tools at the frontiers of quantitative verification. CoRR  \textbf{abs/2405.13583} (2024)

\bibitem{DBLP:conf/cav/AndriushchenkoBCJKM23}
Andriushchenko, R., Bork, A., Ceska, M., Junges, S., Katoen, J., Mac{\'{a}}k, F.: Search and explore: Symbiotic policy synthesis in pomdps. In: {CAV} {(3)}. Lecture Notes in Computer Science, vol. 13966, pp. 113--135. Springer (2023)

\bibitem{DBLP:conf/uai/Andriushchenko022}
Andriushchenko, R., Ceska, M., Junges, S., Katoen, J.: Inductive synthesis of finite-state controllers for pomdps. In: {UAI}. Proceedings of Machine Learning Research, vol.~180, pp. 85--95. {PMLR} (2022)

\bibitem{andriushchenkoPAYNTToolInductive2021}
Andriushchenko, R., Ceska, M., Junges, S., Katoen, J., Stupinsk{\'{y}}, S.: {PAYNT:} {A} tool for inductive synthesis of probabilistic programs. In: {CAV} {(1)}. Lecture Notes in Computer Science, vol. 12759, pp. 856--869. Springer (2021)

\bibitem{DBLP:journals/jair/AndriushchenkoCMJK25}
Andriushchenko, R., Ceska, M., Mac{\'{a}}k, F., Junges, S., Katoen, J.: An oracle-guided approach to constrained policy synthesis under uncertainty. J. Artif. Intell. Res.  \textbf{82},  433--469 (2025)

\bibitem{DBLP:journals/iandc/Angluin87}
Angluin, D.: Learning regular sets from queries and counterexamples. Inf. Comput.  \textbf{75}(2),  87--106 (1987)

\bibitem{babaeeAcceleratedLearningPredictive2019}
Babaee, R., Ganesh, V., Sedwards, S.: Accelerated learning of predictive runtime monitors for rare failure. In: {RV}. Lecture Notes in Computer Science, vol. 11757, pp. 111--128. Springer (2019)

\bibitem{babaeeEmphPreventPredictiveRunTime2018}
Babaee, R., Gurfinkel, A., Fischmeister, S.: \emph{P}revent : {A} predictive run-time verification framework using statistical learning. In: {SEFM}. Lecture Notes in Computer Science, vol. 10886, pp. 205--220. Springer (2018)

\bibitem{DBLP:conf/tacas/BadingsVJSJ24}
Badings, T.S., Volk, M., Junges, S., Stoelinga, M., Jansen, N.: Ctmcs with imprecisely timed observations. In: {TACAS} {(2)}. Lecture Notes in Computer Science, vol. 14571, pp. 258--278. Springer (2024)

\bibitem{baierPrinciplesModelChecking2008}
Baier, C., Katoen, J.: Principles of model checking. {MIT} Press (2008)

\bibitem{baierComputingConditionalProbabilities2014}
Baier, C., Klein, J., Kl{\"{u}}ppelholz, S., M{\"{a}}rcker, S.: Computing conditional probabilities in markovian models efficiently. In: {TACAS}. Lecture Notes in Computer Science, vol.~8413, pp. 515--530. Springer (2014)

\bibitem{DBLP:journals/corr/abs-2412-11754}
Baier, C., Kl{\"{u}}ppelholz, S., Piribauer, J., Ziemek, R.: Formal quality measures for predictors in markov decision processes. CoRR  \textbf{abs/2412.11754} (2024)

\bibitem{bartocciLecturesRuntimeVerification2018}
Bartocci, E., Falcone, Y. (eds.): Lectures on Runtime Verification - Introductory and Advanced Topics, Lecture Notes in Computer Science, vol. 10457. Springer (2018)

\bibitem{DBLP:conf/tacas/BloemKKW15}
Bloem, R., K{\"{o}}nighofer, B., K{\"{o}}nighofer, R., Wang, C.: Shield synthesis: Runtime enforcement for reactive systems. In: International Conference on Tools and Algorithms for the Construction and Analysis of Systems ({TACAS}). Lecture Notes in Computer Science, vol.~9035, pp. 533--548. Springer (2015)

\bibitem{cairoliNeuralPredictiveMonitoring2021}
Cairoli, F., Bortolussi, L., Paoletti, N.: Neural predictive monitoring under partial observability. In: {RV}. Lecture Notes in Computer Science, vol. 12974, pp. 121--141. Springer (2021)

\bibitem{DBLP:conf/aaai/Carr0JT23}
Carr, S., Jansen, N., Junges, S., Topcu, U.: Safe reinforcement learning via shielding under partial observability. In: {AAAI}. pp. 14748--14756. {AAAI} Press (2023)

\bibitem{DBLP:conf/aaai/ChatterjeeCD16}
Chatterjee, K., Chmelik, M., Davies, J.: A symbolic sat-based algorithm for almost-sure reachability with small strategies in pomdps. In: {AAAI}. pp. 3225--3232. {AAAI} Press (2016)

\bibitem{cleavelandConservativeSafetyMonitors2023}
Cleaveland, M., Sokolsky, O., Lee, I., Ruchkin, I.: Conservative safety monitors of stochastic dynamical systems. In: {NFM}. Lecture Notes in Computer Science, vol. 13903, pp. 140--156. Springer (2023)

\bibitem{DBLP:journals/sttt/FalconeFM12}
Falcone, Y., Fernandez, J., Mounier, L.: What can you verify and enforce at runtime? International Journal on Software Tools for Technology Transfer  \textbf{14}(3),  349--382 (2012)

\bibitem{DBLP:conf/aaai/FultonP18}
Fulton, N., Platzer, A.: Safe reinforcement learning via formal methods: Toward safe control through proof and learning. In: {AAAI}. {AAAI} Press (2018)

\bibitem{groceAdaptiveModelChecking2002}
Groce, A., Peled, D.A., Yannakakis, M.: Adaptive model checking. Log. J. {IGPL}  \textbf{14}(5),  729--744 (2006)

\bibitem{DBLP:conf/tacas/HartmannsJQW23}
Hartmanns, A., Junges, S., Quatmann, T., Weininger, M.: A practitioner's guide to {MDP} model checking algorithms. In: {TACAS} {(1)}. Lecture Notes in Computer Science, vol. 13993, pp. 469--488. Springer (2023)

\bibitem{DBLP:journals/jacm/Hastad01}
H{\aa}stad, J.: Some optimal inapproximability results. J. {ACM}  \textbf{48}(4),  798--859 (2001)

\bibitem{DBLP:series/lncs/HavelundRR19}
Havelund, K., Reger, G., Rosu, G.: Runtime verification past experiences and future projections. In: Computing and Software Science, Lecture Notes in Computer Science, vol. 10000, pp. 532--562. Springer (2019)

\bibitem{henselProbabilisticModelChecker2022}
Hensel, C., Junges, S., Katoen, J., Quatmann, T., Volk, M.: The probabilistic model checker storm. Int. J. Softw. Tools Technol. Transf.  \textbf{24}(4),  589--610 (2022)

\bibitem{jansen-et-al-concur-2020}
Jansen, N., K{\"{o}}nighofer, B., Junges, S., Serban, A., Bloem, R.: Safe reinforcement learning using probabilistic shields (invited paper). In: International Conference on Concurrency Theory ({CONCUR}). LIPIcs, vol.~171, pp. 3:1--3:16. Schloss Dagstuhl - Leibniz-Zentrum f{\"{u}}r Informatik (2020)

\bibitem{DBLP:journals/acta/JhaS17}
Jha, S., Seshia, S.A.: A theory of formal synthesis via inductive learning. Acta Informatica  \textbf{54}(7),  693--726 (2017)

\bibitem{jungesActiveLearningRuntime2025}
Junges, S., Seshia, S.A., Torfah, H.: Active learning of runtime monitors under uncertainty. In: {IFM}. Lecture Notes in Computer Science, vol. 15234, pp. 297--306. Springer (2024)

\bibitem{jungesRuntimeMonitorsMarkov2021}
Junges, S., Torfah, H., Seshia, S.A.: Runtime monitors for markov decision processes. In: {CAV} {(2)}. Lecture Notes in Computer Science, vol. 12760, pp. 553--576. Springer (2021)

\bibitem{muskardinAALpyActiveAutomata2022}
Muskardin, E., Aichernig, B.K., Pill, I., Pferscher, A., Tappler, M.: Aalpy: an active automata learning library. Innov. Syst. Softw. Eng.  \textbf{18}(3),  417--426 (2022)

\bibitem{DBLP:journals/jalc/PeledVY02}
Peled, D.A., Vardi, M.Y., Yannakakis, M.: Black box checking. J. Autom. Lang. Comb.  \textbf{7}(2),  225--246 (2002)

\bibitem{DBLP:conf/nfm/PhanG0PSS20}
Phan, D.T., Grosu, R., Jansen, N., Paoletti, N., Smolka, S.A., Stoller, S.D.: Neural simplex architecture. In: {NFM}. Lecture Notes in Computer Science, vol. 12229, pp. 97--114. Springer (2020)

\bibitem{rabinerTutorialHiddenMarkov1989}
Rabiner, L.R.: A tutorial on hidden markov models and selected applications in speech recognition. Proc. {IEEE}  \textbf{77}(2),  257--286 (1989)

\bibitem{RW87}
Ramadge, P.J., Wonham, W.M.: Supervisory control of a class of discrete event processes. SIAM Journal on Control and Optimization  \textbf{25}(1),  206--230 (1987). \doi{10.1137/0325013}

\bibitem{sanchezSurveyChallengesRuntime2019}
S{\'{a}}nchez, C., Schneider, G., Ahrendt, W., Bartocci, E., Bianculli, D., Colombo, C., Falcone, Y., Francalanza, A., Krstic, S., Louren{\c{c}}o, J.M., Nickovic, D., Pace, G.J., Rufino, J., Signoles, J., Traytel, D., Weiss, A.: A survey of challenges for runtime verification from advanced application domains (beyond software). Formal Methods Syst. Des.  \textbf{54}(3),  279--335 (2019)

\bibitem{sistlaMonitoringTemporalProperties2008}
Sistla, A.P., Srinivas, A.R.: Monitoring temporal properties of stochastic systems. In: {VMCAI}. Lecture Notes in Computer Science, vol.~4905, pp. 294--308. Springer (2008)

\bibitem{sistlaRuntimeMonitoringStochastic2012}
Sistla, A.P., Zefran, M., Feng, Y.: Runtime monitoring of stochastic cyber-physical systems with hybrid state. In: {RV}. Lecture Notes in Computer Science, vol.~7186, pp. 276--293. Springer (2011)

\bibitem{smeenkApplyingAutomataLearning2015}
Smeenk, W., Moerman, J., Vaandrager, F.W., Jansen, D.N.: Applying automata learning to embedded control software. In: {ICFEM}. Lecture Notes in Computer Science, vol.~9407, pp. 67--83. Springer (2015)

\bibitem{stollerRuntimeVerificationState2012}
Stoller, S.D., Bartocci, E., Seyster, J., Grosu, R., Havelund, K., Smolka, S.A., Zadok, E.: Runtime verification with state estimation. In: {RV}. Lecture Notes in Computer Science, vol.~7186, pp. 193--207. Springer (2011)

\bibitem{DBLP:conf/atva/TorfahXJVS22}
Torfah, H., Xie, C., Junges, S., Vazquez{-}Chanlatte, M., Seshia, S.A.: Learning monitorable operational design domains for assured autonomy. In: {ATVA}. Lecture Notes in Computer Science, vol. 13505, pp. 3--22. Springer (2022)

\bibitem{DBLP:journals/cacm/Vaandrager17}
Vaandrager, F.W.: Model learning. Commun. {ACM}  \textbf{60}(2),  86--95 (2017)

\bibitem{wilcoxRuntimeVerificationStochastic2010}
Wilcox, C.M., Williams, B.C.: Runtime verification of stochastic, faulty systems. In: {RV}. Lecture Notes in Computer Science, vol.~6418, pp. 452--459. Springer (2010)

\bibitem{yangImprovingModelInference2019}
Yang, N., Aslam, K., Schiffelers, R.R.H., Lensink, L., Hendriks, D., Cleophas, L., Serebrenik, A.: Improving model inference in industry by combining active and passive learning. In: {SANER}. pp. 253--263. {IEEE} (2019)

\bibitem{zolfagharianSMARLASafetyMonitoring2025}
Zolfagharian, A., Abdellatif, M., Briand, L.C., S., R.: {SMARLA:} {A} safety monitoring approach for deep reinforcement learning agents. CoRR  \textbf{abs/2308.02594} (2023)

\end{thebibliography}
\clearpage

\appcite{}{
\appendix
\crefalias{section}{appendix}
\crefalias{subsection}{appsec}

\section{Proof Outlines for \cref{sec:monver}}\label{app:proofs}
\toctp*
\begin{proofoutline}
    By the product construction, there exists a bijective mapping between paths in $M_{\times\overline{\dfa}}$ and paths in the monitor $\dfa$ and HMM $\mc$. If the final state of a path is in $T$, the mapped path is not accepted by the monitor. The risk of a path in $M_{\times\overline{\dfa}}$ has the same risk as the mapped path in $\mc$. Thus, if a trace is a witness for the CTR problem, it is a missed alarm for $\mc$ and $\dfa$.
\end{proofoutline}

\toactp*
\begin{proofoutline}
    We choose $\lambda = \sfrac{\lambda_u}{\max_{s\in S}r(s)}$. Both directions follow from applying \cref{def:CTPdecprob} and defining a bijective mapping $f$ between paths of length the horizon in $\mc_{\times\overline{\dfa}}$ (From the CTR problem) and $\mc_{\blacktriangleright h}$ such that $\Prob^{\mc_{\times\overline{\dfa}}}(\pi) = \Prob^{\mc_{\blacktriangleright h}}(f(\pi))$. Now, for any path $\pi$ of length $h$ ending in a state in $T$, the transition of $f(\pi_\downarrow)$ to $\talrm$ is equal to the normalized risk on $\pi_\downarrow$.
    Thus, any trace which is a witness of CTR has a summed probability above $\lambda$ in $\mc_{\blacktriangleright h}$.
\end{proofoutline}

\tocoloredmdp*
\begin{proofoutline}
    Given a trace $\tau$ we define a bijective map $f_\tau$ between finite paths $\pi$ in $\mc_{\blacktriangleright h}$ with $\Prob^{\mc_{\blacktriangleright h}}(\tau\mid\pi) = 1$, and a set $X$ in the partition of the infinite paths in the induced MC by $\sigma_\tau$ in $\mdp_{\gtrdot h}$. $f$ is defined such that a path $\pi$ maps to the set of paths $\{\pi'\cdot\pi\cdot\alarms^\star\mid\pi'\in\Pi^{\mdp_{\gtrdot h}^\sigma}\}$. Using \cref{def:polsyntrans}, we can prove that the probability of $\pi$ and $X$ are equal. Now, using a bijective map $t$ between traces and trace consistent policies, we can show that $\Prob^{\mdp_{\gtrdot h}}_{t(\tau)}(\lozenge T) = \sum_{\pi\in\Pi^{\mc_{\blacktriangleright h}}}\Prob^{\mc_{\blacktriangleright h}}(\pi \cdot \talrm \mid \tau \cdot \en)$. Thus, if there exists a trace consistent policy $\sigma_\tau$ above the threshold, $t(\sigma_\tau)$ is also above the threshold.
    The other direction follows similarly using $t^{-1}$ and assuming a policy.
\end{proofoutline}

\leqhtheorem*
\begin{proof}
    We modify the transformation from \cref{thm:maintheorem} in the policy synthesis step. We add a new initial state, $\iota'$, which gets a separate coloring from all other states. This state contains an action for each possible length $l$ of a trace up to the horizon. Taking action $l$ leads to the state $\langle l, \iota\rangle$.
    \begin{align*}
        \ptrans'(\iota', l)(\langle l, \iota\rangle) & = 1
    \end{align*}
    We now show this transformation is correct.

    In order to verify that there are no-missed-alarms for all $l<h$, we could use the transformation from \cref{thm:maintheorem} with the horizon equal to all $l<h$. This would necessitate doing policy synthesis for $h$ colored MDPs.
    \[ \exists_{\tau} \exists_{l\leq h} \Prob^{\sigma_\tau}[\lozenge alarm] > \lambda_u \]
    If any of these $h$ policy synthesis problems can find a policy $\sigma_\tau$, there exists a trace $\tau\in\missedalarms{\leq h}{}$.

    We combine these $h$ colored MDPs into one colored MDP in the following way.
    We add a new initial state, and give it $h$ actions $\{1,\ldots,h\}$. Action $l\in\{1,\ldots,h\}$ points, with probability $1$, to the initial state of colored MDP $\mdp_{\gtrdot l}$. This is equivalent to solving policy synthesis on the $h$ individual MDPs.

    We now note that for any $l<l'\leq h$ the initial state $\langle 1, \iota^{CTR}\rangle$ of $\mdp_{\gtrdot l}$ is bisimilar to the state $\langle l' - l + 1, \iota^{CTR} \rangle$ in $\mdp_{\gtrdot l'}$. We now claim that the result of bisimulation minimization on this combined colored MDP is described by the transformation described at the start of the proof.
\end{proof}

\nfa*
\begin{proofoutline}
    The transformation from \cref{sec:monver} is reused with the following differences. The complement of the monitor in \cref{lem:toctp} is no longer taken, and $\safes$ is used as the target state in \cref{lem:tocoloredmdp}. An outline of why the second step is correct is given below by showing the following:

    \[
        \begin{array}{c}
            \exists \sigma \in \ccSched.  \Prob^{\mdp_{\gtrdot h}}_{\sigma}(\lozenge \safes) > 1 - \lambda \\
            \Updownarrow                                                                                   \\
            \exists \tau \in \lang{\mc_{\blacktriangleright h}}. \sum_{\pi\in\Pi^{\mc_{\blacktriangleright h}}}\Prob^{\mc_{\blacktriangleright h}}(\pi \cdot \talrm \mid \tau \cdot \en) \leq \lambda
        \end{array}
    \]
    Using the proof of \cref{lem:tocoloredmdp} where we replace $\talrm$ with $\tsafe$, and $\lambda$ with $1 - \lambda$, results in the following statement:
    \[
        \begin{array}{c}
            \exists \tau \in \lang{\mc_{\blacktriangleright h}}. \sum_{\pi\in\Pi^{\mc_{\blacktriangleright h}}}\Prob^{\mc_{\blacktriangleright h}}(\pi \cdot \tsafe \mid \tau \cdot \en) > 1 - \lambda \\
            \Updownarrow                                                                                                                                                                               \\
            \exists \tau \in \lang{\mc_{\blacktriangleright h}}. \sum_{\pi\in\Pi^{\mc_{\blacktriangleright h}}}\Prob^{\mc_{\blacktriangleright h}}(\pi \cdot \talrm \mid \tau \cdot \en) \leq \lambda
        \end{array}
    \]
    This can be shown to hold using the following fact,
    \begin{align*}
        \forall\tau \in \lang{\mc_{\blacktriangleright h}} \sum_{\pi\in\Pi^{\mc_{\blacktriangleright h}}}
        \left(
        \begin{array}{l}
                \Prob^{\mc_{\blacktriangleright h}}(\pi \cdot \tsafe \mid \tau \cdot \en)\ + \\
                \Prob^{\mc_{\blacktriangleright h}}(\pi \cdot \talrm \mid \tau \cdot \en)
            \end{array}
        \right)
        \in \{0,1\}
    \end{align*}
\end{proofoutline}

\learncor*
\begin{proof}
    If, while learning a monitor $\dfa$ a trace $\tau$ is deemed \emph{safe} by $\MQ_{\lambda_l}$ it cannot be given as a counterexample by $\EQ_{\lambda_s, \lambda_u}$ on $\dfa$, since $\lambda_s \leq \lambda_l \leq \lambda_u$, and $\tau\notin\dfa$ by L$^\star$. Similarly, $\MQ_{\lambda_l}$ and $\EQ_{\lambda_s, \lambda_u}$ also agree on \emph{unsafe} traces. Now, by correctness of L$^\star$, a learned monitor $\dfa$ has to be correct according to $\EQ_{\lambda_s, \lambda_u}$, and thus correct in the sense of \cref{def:cormon}.
\end{proof}

\section{Construction for NP-hardness/APX-hardness}
\label{app:complexity}

Consider a 3CNF formula $\varphi = \bigwedge c_1 \dots c_m$ over variables $X$, $|X|=n$, with clause $c_j = \ell^1_j \lor \ell^2_j \lor \ell^3_j$ and each literal $\ell^i_j \in \{ x, \neg x \mid x \in X\}$. We transform this into a CTR instance with $\lambda = 1$ and an acyclic HMM $\mc_\varphi$  with observations $\Obs= \{ \#, \bot, \top \}$. The only state with positive risk  is a dedicated state $t$ with $r(t) = 1$, we also set $T=\{t\}$. The crux of the construction is that there is a trace with risk $1$ iff there is a satisfying assignment to $\varphi$. In the constructed HMM, there is a trace with risk $1$ iff there is a trace where all corresponding paths end in state $t$.

Before we give a formal definition of $\mc_\varphi$, we give some intuition.
We represent assignments $\alpha\colon X \rightarrow \{\bot,\top \}$ by traces through the HMM $\#\# \cdot \alpha(x_0) \cdot \# \cdot \alpha(x_1) \dots \cdot \#$. We create gadgets for every clause. The gadgets are connected as in \cref{fig:gadgetsconnected}: That is, intuitively, $\mc_\varphi$ randomly selects a clause $c_j$ with probability $\nicefrac{1}{m}$ and transitions into gadget $\mathcal{G}_j$ defined below. Next, we show that the gadget reaches positive risk in every gadget only if the corresponding clause is satisfied by the assignment.

The gadget $\mathcal{G}_j$ intuitively `evaluates' $c_j$ with respect to an assignment $\alpha$, by starting in $s_{1,j}$ with a gadget for this clause, as exemplified in Fig.~\ref{fig:hmmgadget}.  We enter the gadget with trace $\# \cdot \alpha(x_0) \cdot \# \cdot \alpha(x_1) \dots \cdot \#$, i.e., the first observation has been matched by the initial state.
A path (or its trace) through $\mathcal{G}_j$ `reads' variable $x_i$ in state $s_{i,j}$ and transitions to $s_{i,j}^\top$ or to $s_{i,j}^\bot$. However, only one of these paths corresponds to the trace. Thus, in every gadget, there is only one path that corresponds to a given trace and so they can be used interchangeably.
State $s_{i,j}$, $s_{i,j}^\top$ and $s_{i,j}^\bot$ have observations $\#, \top, \bot$, respectively. We note that a path  reaches $s_{i,j}^\top$ if $\alpha(x_i) = \top$ and (analogously for $\bot$).
From there onwards,  if the literal $x_i$ (or $\neg x_i$) occurs in clause $c_j$, we transition from $s_{i,j}^\top$ (or $s_{i,j}^\bot$, resp.) to $s_{i+1,j}'$, otherwise, we transition to $s_{i+1,j}$. A path ends in $t$ only via a state $s'_{i,j}$, which is only possible if the path visits a state corresponding to a literal in the clause. Thus, (conditionally) reaching state $s'_{i,j}$ with positive conditional probability means that the partial assignment $\alpha(x_1),\dots,\alpha(x_i)$ satisfies clause $c_j$, while reaching $s_{i,j}$ with positive conditional probability means that the clause is either unresolved or unsatisfied given the partial assignment.

Formally, we construct the HMM $\mc = (\Sts,\init, \ptrans, \Obs, \obsf, r)$ and set $\lambda_s = 1$, where we use $[\varphi]$ to be the indicator function of $\varphi$:
\begin{itemize}
    \item
          $\Sts = \{ s_{i,j}, s_{i,j}^\bot, s_{i,j}^\top,
              s_{i,j}', s_{i,j}'^\bot, s_{i,j}'^\top
              \mid i \in \{1, \dots, n+1 \},  j \in \{ 1, \dots, m \} \} \cup \{ s_\init \}$.
    \item $\ptrans$ is given such that for all $i \in \{1,\dots, n\}, j \in \{1, \dots, m\}$:
          \begin{itemize}
              \item  $\ptrans(s_\init,s_{1,j}) = \frac{1}{m}$,
              \item  $\ptrans(s_{i,j},s_{i,j}^\top) = \ptrans(s_{i,j},s_{i,j}^\bot) = \frac{1}{2}= \ptrans(s'_{i,j},s_{i,j}'^\top) = \ptrans(s_{i,j}',s_{i,j}'^
                        \bot)$,
              \item $\ptrans(s_{i,j}^\top,s'_{i+1,j}) = \indicator{x_i = \ell_j^k \text{ for some k}}$, $\ptrans(s_{i,j}^\top,s_{i+1,j}) = \indicator{x_i \neq \ell_j^k \text{ for all k}}$,
              \item $\ptrans(s_{i,j}^\bot,s'_{i+1,j}) = \indicator{\neg x_i = \ell_j^k \text{ for some k}}$, $\ptrans(s_{i,j}^\bot,s_{i+1,j}) = \indicator{\neg x_i \neq \ell_j^k \text{ for all k}}$
              \item $\ptrans(s'_{n+1,j}, t) = 1 = \ptrans(s_{n+1,j}, f)$
          \end{itemize}
    \item $\Obs = \{\top, \bot, \#\}$ and $\obsf(s_{i,j}^\top) = (s_{i,j}'^\top) = \top$,
          $\obsf(s_{i,j}^\bot) = (s_{i,j}'^\bot) = \bot$,
          $\obsf(s_{i,j}) = \obsf(s'_{i,j}) = \# = \obsf(s_\init) = \obsf(t) = \obsf(f)$.
    \item $r(t) = 1$ and $r(s) = 0$ for all $s \in \Sts \setminus \{ t\}$.
\end{itemize}
The construction runs in polynomial time. The formal proof of its correctness follows the explanation above precisely.

\section{Results from \alg Verification Experiments} \label{sec:resverify}
\subsection{Transformation Time Results}\label{app:transtimeres}
We present the complete results of the monitor verification experiments. \cref{fig:sizevsprod} (log scale!) shows how transforming the HMM with the monitor into a colored MDP scales with the number of states in each.
\subsection{Verification Results Table}\label{app:verrestable}
\Cref{tab:fullverifyres} contains the full results of the monitor verification experiments. The columns give the family name, an ID, learning threshold, horizon, and whether we check for missed alarms or false alarms. We give the size of the HMM (states, transitions), the number of observations, the size of the DFA (states, transitions), and the approximate size of the language of the HMM. Furthermore, we list the run time for the complete verification procedure as well as the time spent on transforming the problem into a policy synthesis problem and the policy synthesis in PAYNT. We list the size of the created Colored MDP, and finally the found threshold. Checkmarks in $\lambda_l$ indicate that, instead of verifying a bound, we are verifying an extremum. Checkmarks in $\lambda^{found}$ indicate no trace was found with a risk above (or below, for FA) the indicated threshold. If all values in the \alg section of a row contain dashes, either a time-out or out-of-memory occurred, or the target state was not reachable because of the horizon.

\begin{figure}[t]
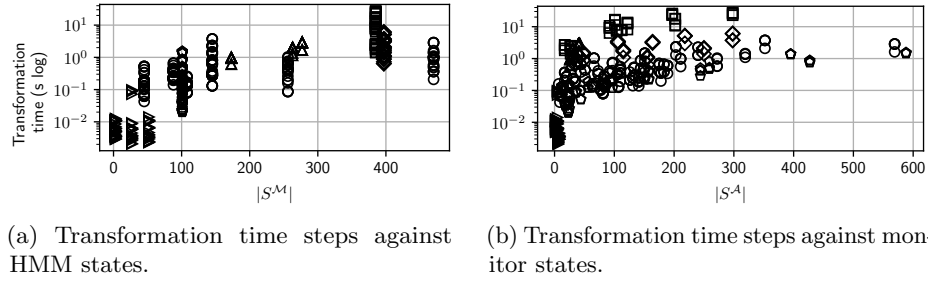

    \begin{subfigure}{0.48\linewidth}
        \begin{adjustbox}{max width=\linewidth, clip, trim=.2cm .2cm .2cm .2cm}
            \input{images/plots/hmm_states_vs_product_runtime.pgf}
        \end{adjustbox}
        \caption{Transformation time steps against HMM states.}
    \end{subfigure}\hfill
    \begin{subfigure}{0.48\linewidth}
        \begin{adjustbox}{max width=\linewidth, clip, trim=.2cm .2cm .2cm .2cm}
            \input{images/plots/mon_states_vs_product_runtime.pgf}
        \end{adjustbox}
        \caption{Transformation time steps against monitor states.}
    \end{subfigure}
    \caption{Time in the transformation step compared to the size of the HMM $\mc$ or the monitor $\dfa$ for \alg verification.}
    \label{fig:sizevsprod}
\end{figure}

\begin{scriptsize}
    \setlength\LTleft{-1.2cm}
\begin{longtable}[c]{@{}llrrrrrrrrrrrrrr@{}}
\caption{Table of all verification experiments.}
\label{tab:fullverifyres}\\                                                                                                                                                                                                                                                                                                        \\
\toprule
 & & \multicolumn{8}{c}{Benchmark} & \multicolumn{5}{c}{\alg}                                                                                                                                                       \\
\cmidrule(lr){3-11}\cmidrule(lr){12-16}
 & & $\lambda_l$ & $h$ & MA/FA & $|\Sts^\mc|$ & $|\ptrans^\mc|$ & $|Z|$ & $|\Sts^\dfa|$ & $|\ptrans^\dfa|$ & $|\lang{\mc}^{\leq h}|$ & Time (s) & Trans (s) & PAYNT (s) & $|\mdp_{\gtrdot h}|$ & $\lambda^{found}$  \\
\midrule
\endhead\textsc{AirportA-3} & \textsc{A-0} & \sfrac{3}{10} & 10 & MA & 45 & 113 & 18 & 34 & 612 & $10^{13}$ & $\leq 1s$ & $\leq 1s$ & $\leq 1s$ & 383 & 1.00\\
\textsc{AirportA-3} & \textsc{A-1} & \checkmark & 10 & MA & 45 & 113 & 18 & 34 & 612 & $10^{13}$ & $\leq 1s$ & $\leq 1s$ & $\leq 1s$ & 383 & 1.00\\
\textsc{AirportA-3} & \textsc{A-2} & \sfrac{3}{10} & 10 & FA & 45 & 113 & 18 & 34 & 612 & $10^{13}$ & $\leq 1s$ & $\leq 1s$ & $\leq 1s$ & 383 & 0.09\\
\textsc{AirportA-3} & \textsc{A-3} & \checkmark & 10 & FA & 45 & 113 & 18 & 34 & 612 & $10^{13}$ & 294 & $\leq 1s$ & 293 & 383 & 0.07\\
\textsc{AirportA-3} & \textsc{A-4} & \sfrac{3}{10} & 10 & MA & 45 & 113 & 18 & 73 & 1314 & $10^{13}$ & 261 & $\leq 1s$ & 261 & 565 & 0.45\\
\textsc{AirportA-3} & \textsc{A-5} & \checkmark & 10 & MA & 45 & 113 & 18 & 73 & 1314 & $10^{13}$ & 798 & $\leq 1s$ & 798 & 565 & 1.00\\
\textsc{AirportA-3} & \textsc{A-6} & \sfrac{3}{10} & 10 & FA & 45 & 113 & 18 & 73 & 1314 & $10^{13}$ & 1 & $\leq 1s$ & $\leq 1s$ & 565 & 0.25\\
\textsc{AirportA-3} & \textsc{A-7} & \checkmark & 10 & FA & 45 & 113 & 18 & 73 & 1314 & $10^{13}$ & 588 & $\leq 1s$ & 588 & 565 & 0.07\\
\bottomrule
\end{longtable}
            
\end{scriptsize}

\section{Results from Baseline Sampling Count Experiment}\label{sec:ressampling}
We evaluate the impact of the learning threshold $\lambda_l$ and the amount of samples used during conformance testing for the baseline model. \cref{fig:samplingn} contains the minimum risk of a trace accepted by the monitor and the maximum risk of a trace not accepted by the monitor for
$\lambda_l \in \{0.05, 0.2, 0.4\}$ and sampling counts in $\{100,1000,10000,100000\}$. All combinations where evaluated on \textsc{AiportB-3} and \textsc{AirportB-7}.

Using $10000$ samples and $100000$ samples both had the same number of missed alarms. However, $10000$ Samples had more false alarms then $100000$ samples. The learning threshold seemed to have no effect on the correctness of the learned monitors.
\begin{figure}[h]
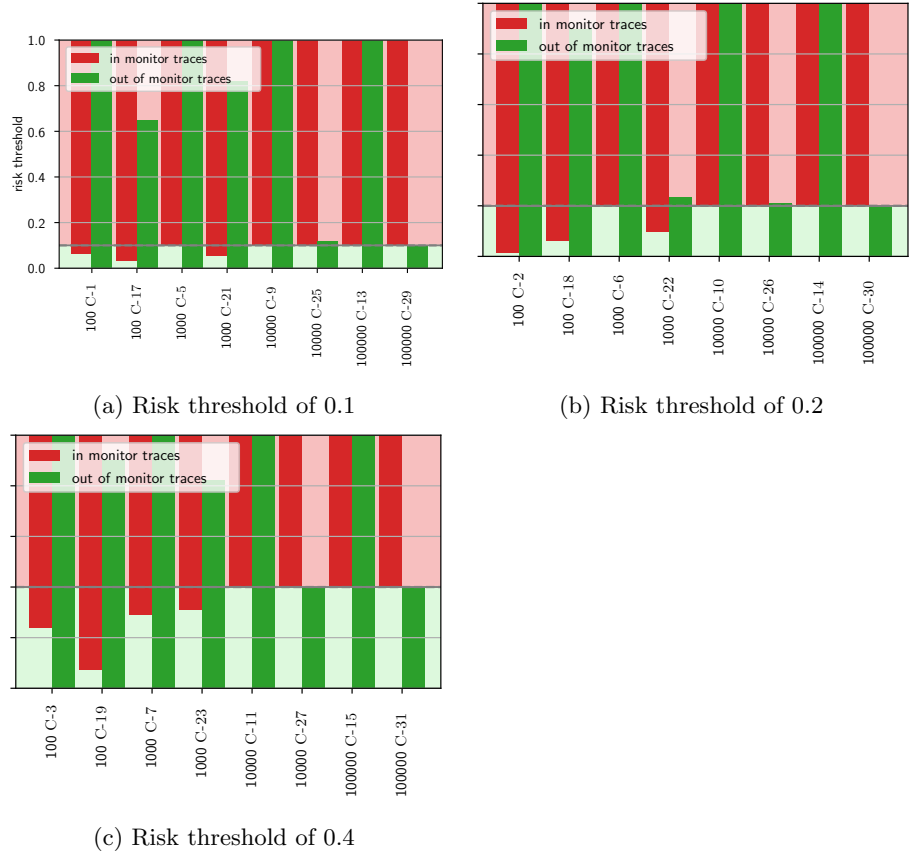

    \begin{subfigure}{.5\linewidth}
        \resizebox{\linewidth}{!}{
            \input{images/plots/sampling_thresholds_low.pgf}
        }
        \caption{Risk threshold of 0.1}
    \end{subfigure}
    \begin{subfigure}{.5\linewidth}
        \resizebox{\linewidth}{!}{
            \input{images/plots/sampling_thresholds_mid.pgf}
        }
        \caption{Risk threshold of 0.2}
    \end{subfigure}
    \begin{subfigure}{.5\linewidth}
        \resizebox{\linewidth}{!}{
            \input{images/plots/sampling_thresholds_high.pgf}
        }
        \caption{Risk threshold of 0.4}
    \end{subfigure}
    \caption{Found risk thresholds when using baseline leaning with a different amount of samples in each EQ step. This experiment is done for three different risk thresholds.}
    \label{fig:samplingn}
\end{figure}

\section{Results from Learning Experiments}\label{sec:reslearn}

We present the complete results for the Learning experiments.

\subsection{Learning Figures Legend}\label{app:learnlegend}
\cref{fig:legend} contains the symbol legend for all learning and verification experiment figures.

\begin{figure}
    \centerline{
        \resizebox{.5\linewidth}{!}{
            \input{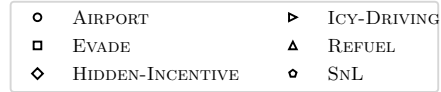}}
    }
    \caption{Legend of symbols used in plots}
    \label{fig:legend}
\end{figure}

\subsection{Results Table of Learning Experiment}\label{app:learnrestable}
\Cref{tab:fulllearnexp} contains the full results of the learning experiments. The columns give the family name and ID. They also list the threshold parameters and the horizon. We report the size of the HMM (states, transitions, number of observations). Furthermore, we detail the results of \alg learning. We give the total runtime for the learning procedure. We also show the amount of EQs needed and the number of states in the learned monitor. Additionally, we present the found minimum risk of a trace accepted by the monitor and the maximum risk of a trace not accepted by the monitor. Lastly, we detail the results of the baseline learning method. We again list the amount of time spent, number of EQs, the number of states in the learned monitor, and the minimum and maximum threshold as before. Any red bounds on lambda indicate incorrect monitors.

Whenever either the \alg columns of a row or the baseline columns of a row contain dashes, its time column gives the reason for not learning a monitor. $\infty$ means the 12 hour time-out was exceeded. OM means the 15 GiB memory limit was exceeded.

\begin{scriptsize}
    \setlength\LTleft{-.2cm}
\begin{longtable}[c]{@{}llrrrrrrrrrrrrrrrr@{}}
\caption{Table of all learn experiments.}
\label{tab:fulllearnexp}\\                                                                                                                                                                                                                                                                                                    \\
\toprule
 & & \multicolumn{6}{c}{Benchmark} & \multicolumn{5}{c}{\alg} & \multicolumn{5}{c}{Baseline}                                                                                                                                                       \\
\cmidrule(lr){3-8}\cmidrule(lr){9-13}\cmidrule(lr){14-18}
 & & $\lambda_s$ & $\lambda_u$ & $h$ & $|\Sts|$ & $|\ptrans|$ & $|Z|$ & Time (s) & Rounds & $|\dfa|$ & $\lambda_u^{\min}$ & $\lambda_s^{\max}$ & Time (s) & $|\dfa|$ & Rounds & $\lambda_u^{\min}$ & $\lambda_s^{\max}$ \\
\midrule
\endhead\textsc{Airport} & \textsc{A-0} & \sfrac{3}{10} & \sfrac{3}{10} & 10 & 45 & 113 & 18 & 11425 & 26 & 133 & 0.30 & 0.30 & 148 & 196 & 39 & 0.30 & 0.30\\
\textsc{Airport} & \textsc{A-1} & \sfrac{1}{10} & \sfrac{7}{20} & 10 & 45 & 113 & 18 & 10767 & 22 & 101 & 0.12 & 0.32 & 148 & 196 & 39 & 0.30 & 0.30\\
\textsc{Airport} & \textsc{A-2} & \sfrac{3}{10} & \sfrac{3}{10} & 11 & 45 & 113 & 18 & $\infty$ & - & - & - & - & 271 & 262 & 46 & 0.30 & 0.30\\
\textsc{Airport} & \textsc{A-3} & \sfrac{1}{10} & \sfrac{7}{20} & 11 & 45 & 113 & 18 & $\infty$ & - & - & - & - & 271 & 262 & 46 & 0.30 & 0.30\\
\textsc{Airport} & \textsc{A-4} & \sfrac{3}{10} & \sfrac{3}{10} & 8 & 45 & 113 & 18 & 335 & 23 & 107 & 0.31 & 0.30 & 74 & 106 & 31 & 0.31 & 0.30\\
\textsc{Airport} & \textsc{A-5} & \sfrac{1}{10} & \sfrac{7}{20} & 8 & 45 & 113 & 18 & 142 & 10 & 45 & 0.10 & 0.31 & 74 & 106 & 31 & 0.31 & 0.30\\
\textsc{Airport} & \textsc{A-6} & \sfrac{3}{10} & \sfrac{3}{10} & 9 & 45 & 113 & 18 & 2776 & 34 & 141 & 0.31 & 0.30 & 152 & 145 & 37 & 0.31 & 0.30\\
\textsc{Airport} & \textsc{A-7} & \sfrac{1}{10} & \sfrac{7}{20} & 9 & 45 & 113 & 18 & 870 & 14 & 66 & 0.13 & 0.34 & 152 & 145 & 37 & 0.31 & 0.30\\
\bottomrule
\end{longtable}
            
\end{scriptsize}
}

\end{document}